\documentclass{article}

\usepackage{microtype}
\usepackage{graphicx}
\usepackage[font=scriptsize]{subfig}
\usepackage{booktabs} 

\usepackage{mathrsfs,mathtools}
\usepackage{amsmath,amssymb,amsthm}
\usepackage{symbols}

\usepackage{hyperref}

\newcommand{\m}{\mathrm{max}}

\usepackage{xcolor}
\usepackage{enumerate}
\usepackage{dsfont}
\definecolor{DukeBlue}{HTML}{001A57}
\definecolor{DarkRed}{rgb}{0.75, 0.0, 0.0}

\allowdisplaybreaks

\newtheorem{assumption}[theorem]{Assumption}

\usepackage[accepted]{icml2019}

\icmltitlerunning{Multi-Frequency Phase Synchronization}

\begin{document}

\twocolumn[
\icmltitle{Multi-Frequency Phase Synchronization}



\icmlsetsymbol{equal}{*}

\begin{icmlauthorlist}
\icmlauthor{Tingran Gao}{equal,uchicago}
\icmlauthor{Zhizhen Zhao}{equal,uiuc}
\end{icmlauthorlist}

\icmlaffiliation{uchicago}{Committee on Computational and Applied
  Mathematics, Department of Statistics, University of
  Chicago, Chicago IL, USA}
\icmlaffiliation{uiuc}{Department of Electrical and Computer
  Engineering, Coordinated Science Laboratory, University of Illinois at Urbana-Champaign, Urbana IL, USA}

\icmlcorrespondingauthor{Tingran
  Gao}{tingrangao@galton.uchicago.edu}
\icmlcorrespondingauthor{Zhizhen Zhao}{zhizhenz@illinois.edu}

\icmlkeywords{phase synchronization, spectral methods,
  cryo-EM, harmonic retrieval}

\vskip 0.3in
]



\printAffiliationsAndNotice{\icmlEqualContribution} 

\begin{abstract}
We propose a novel formulation
for \emph{phase synchronization}---the statistical problem
of jointly estimating alignment angles from noisy pairwise
comparisons---as a nonconvex 
optimization problem that enforces consistency among the
pairwise comparisons in multiple frequency
channels. Inspired by \emph{harmonic retrieval}  in
signal processing, we develop a simple yet
efficient two-stage algorithm that leverages the
multi-frequency information. We demonstrate in
theory and practice that the proposed algorithm
significantly outperforms state-of-the-art phase
synchronization algorithms, at a mild computational
costs incurred by using the extra frequency channels. We also
extend our algorithmic framework to general synchronization
problems over compact Lie groups.
\end{abstract}

\section{Introduction}
\label{sec:introduction}

\emph{Angular} or \emph{phase synchronization}
\cite{Singer2011,Boumal2016}
concerns estimating angles $\theta_1,\dots,\theta_n$ in $\left[ 0,2\pi
\right)$ from a subset of possibly noise-contaminated relative
offsets $\left(\theta_i-\theta_j\right)\!\!\!\mod 2\pi$. An
instance of phase synchronization can be encoded on an
\emph{observation graph} $G=\left( V,E \right)$, where each
angle is assigned to a vertex $i\in V$ and relative offsets
are measured between 
$\theta_i$ and $\theta_j$ if and only if there is an edge in
$G$ connecting vertices $i$ and $j$. Equivalently, the
angles can be encoded into a column \emph{phase vector} 
$z=(\exp \iota \theta_1,\cdots,\exp\iota \theta_n)^{\top}$, 
and measurements constitute a Hermitian matrix
\begin{equation}
\label{eq:additive-gaussian}
  H=A\circ \left[zz^{*}+\Delta\right],\vspace{-0.02in}
\end{equation}
where $A$ is the adjacency matrix of the observation graph
$G$, $\circ$ is the entrywise product, and the Hermitian
matrix $\Delta\in\mathbb{C}^{n\times n}$ encodes measurement noise.

As a prototypical example of more general
\emph{synchronization problems} arising from many scientific fields
concerning consistent pairwise comparisons within large
collections of objects (e.g., cryogenic electron
microscopy \cite{singer2011viewing} and comparative biology
\cite{GBM2019}), phase synchronization attracted much
attention due to its simple yet rich mathematical
structure. One mathematical formulation  is through
nonconvex optimization \vspace{-0.09in}
\begin{equation}
  \label{eq:opt-prob-ang-sync}
    \max_{x\in \mathbb{C}^n_1} x^{*}Hx \vspace{-0.09in}
\end{equation} 
where $\mathbb{C}_1^n$ is the Cartesian product of $n$
copies of $\Unitary \left( 1 \right)$. Depending on the
context of the scientific problem, $H$ may be assumed to
arise from an \emph{additive Gaussian noise model}
\cite{Boumal2016,BBS2017}, in which the Hermitian matrix
$\Delta$ in \eqref{eq:additive-gaussian} is a Wigner matrix
with i.i.d. complex Gaussian entries above the diagonal, or
from a \emph{random corruption model}
\cite{Singer2011,CSG2016} that assumes
\begin{equation}
  \label{eq:random-corruption-phase}
  H_{ij}=
  \begin{cases}
    z_i\bar{z}_j & \textrm{with prob. $r\in\left[ 0,1 \right]$}\\
    w\sim\mathrm{Unif}\left( \Unitary \left( 1 \right) \right) & \textrm{with prob. $1-r$}
  \end{cases}
\end{equation}
for each edge $\left( i,j \right)\in E$. Note that the
random corruption model can also be cast in the form
\eqref{eq:additive-gaussian} after proper shifting and
scaling. In general, the additive Gaussian noise model is
more amenable to analysis, while the random corruption model
is better at capturing the behavior of physical or imaging
models where many outliers exist.

In this paper, we propose to tackle the phase
synchronization problem by solving an alternative nonconvex
optimization problem of ``multi-frequency'' nature, namely, \vspace{-0.09in}
\begin{equation}
  \label{eq:opt-prob-ang-sync-mult-freq}
    \max_{x\in \mathbb{C}^n_1} \sum_{k=1}^{k_{\mathrm{max}}}(x^{k})^{*}H^{\left( k \right)}x^k \vspace{-0.06in}
\end{equation}
where $k_{\mathrm{max}}$ is the number of frequency channels,
$x^k$ is the entrywise $k$th power of $x$, and $H^{\left( k
  \right)}\in\mathbb{C}^{n\times n}$ is a Hermitian matrix
containing information of the ``true signal'' $z$ in the $k$th
frequency component:
\begin{itemize}
\item For the random corruption model
\eqref{eq:random-corruption-phase}, we construct
$H^{\left( k \right)}$ directly from $H$ by entrywise power:
\begin{equation}
  \label{eq:multi-freq-random-corruption}
  H^{\left( k \right)}_{ij}=H_{ij}^k, \quad
  k=1,\dots,k_{\mathrm{max}}; \,\,1\leq i,j\leq n.
\end{equation}
\item For the additive Gaussian noise model, following \cite{BCS2015,PWBM2018}, we assume
\begin{equation}
  \label{eq:multi-freq-additive-gaussian}
  H^{\left( k \right)}=A\circ [z^k \left( z^{*}
  \right)^k+\sigma_k\Delta^{\left(k\right)}],
\end{equation}
where each $\Delta^{\left(k\right)}$ is a complex Hermitian
random matrix with independent upper diagonal
entries, and the scaling $\sigma_k$ is chosen such that the
operator norm of $\Delta^{\left( k \right)}$ is upper
bounded by $\sqrt{n}$. Unlike
\cite{BCS2015,PWBM2018}, we allow
entries of $\Delta^{\left( k \right)}$ to be general
sub-Gaussian random variables rather than restrictively
complex Gaussian, and we do not assume independence of the
$\Delta^{\left( k \right)}$'s across different $k$'s.
\end{itemize}

We treat the two types of noise
\eqref{eq:multi-freq-random-corruption}
\eqref{eq:multi-freq-additive-gaussian} in a unified 
model, under which we design and analyze our \emph{multi-frequency phase
  synchronization} algorithm. We demonstrate
surprising theoretical and empirical results that
drastically outperform all existing phase 
synchronization algorithms in their corresponding settings,
measured in terms of the correlation between the output
and the true phase vector $z$, at a
mild increase in the computational cost incurred by
parallelizing the computation in $k_{\mathrm{max}}$
frequency channels. As will be demonstrated in
Section~\ref{sec:analysis}, in the noise regime where
phase synchronization is tractable, the number of
frequencies $k_{\mathrm{max}}$ needed to outperform
single frequency algorithms is at
most polylogarithmically dependent on the problem size $n$,
while the estimation error decays polynomially in
$k_{\mathrm{max}}$.

\paragraph{Motivation} The rationale behind the
multi-frequency formulation
\eqref{eq:opt-prob-ang-sync-mult-freq} lies at the
observation that statistical estimation can often benefit
from higher moment estimates, even without introducing new
measurements. As a motivating example, let $G$ be a complete
graph, and consider the following $k_{\mathrm{max}}=2$
coupled problems:
\vspace{-0.06in}
\begin{equation}
\label{eq:coupled-problems}
    \max_{x\in \mathbb{C}^n_1} x^{*}H^{\left( 1 \right)}x; \qquad \max_{x\in \mathbb{C}^n_1} \left(x^{2}\right)^{*}H^{\left( 2 \right)}x^2 \vspace{-0.06in}
\end{equation}
where $H^{\left( 1 \right)}=H$, and $H^{\left( 2 \right)}$
is generated according to
\eqref{eq:multi-freq-random-corruption}. Up to rescaling by
a factor of $1/r$, $H^{\left( 1 \right)}$ and $H^{\left( 2
  \right)}$ fit into model
\eqref{eq:multi-freq-additive-gaussian} with
\begin{equation}
\label{eq:random-corruption-example}
  \sigma_k\Delta_{ij}^{\left( k \right)}=
  \begin{cases}
    \left(1-r\right) z_i\bar{z}_j & \textrm{with prob. $r\in\left[ 0,1 \right]$}\\
    e^{\iota k\varphi_{ij}}-rz_i\bar{z}_j  & \textrm{with prob. $1-r$}
  \end{cases}
\end{equation}
where $\varphi_{ij}$ are i.i.d. uniform on $\mathbb{R}/2\pi$ for $\left( i,j \right)\in E$, and
$\varphi_{ji}=-\varphi_{ij}$. Note that
$\sigma_1\Delta^{\left( 1 \right)}$ and $\sigma_2\Delta^{\left(
    2 \right)}$ are by no means independent, but for all practical
purposes satisfy the same sub-Gaussian bounds since
$e^{\iota\varphi_{ij}}$ and $e^{\iota 2\varphi_{ij}}$ are identically
distributed; we thus assume without loss of generality that
$\sigma_1=\sigma_2$. If we can find
$\hat{x}\in\mathbb{C}_1^n$ satisfying jointly
\begin{align}
\label{eq:separate-max-toy}
(\hat{x}^k)^{*}H^{\left( k \right)}\hat{x}^k \geq (z^k)^*H^{\left( k \right)}z^k\,\,k=1,2
\end{align}
then, by Lemma~1 of \cite{Boumal2016} (assuming
without loss of generality that
$\hat{x}^{*}z=|\hat{x}^{*}z|$), we have for $k=1,2$
\begin{equation}
\label{eq:before-bootstrap}
    \left\| \hat{x}^k-z^k \right\|_2^2=2 ( n-\left|
      z^{*}\hat{x} \right|^k )\leq 16\sigma^2 \| \Delta^{\left( k \right)} \|_2^2/n,
\end{equation}
which gives $\left| z^{*}\hat{x} \right|\geq \max_{k=1,2}
\left\{ \left( n-8\sigma^2\|\Delta^{\left( k
      \right)}\|_2^2/n \right)^{\frac{1}{k}}\right\}$, a
tighter bound than one could obtain from
\eqref{eq:separate-max-toy} with $k=1$ alone, especially for large 
$\sigma$ (with $n-8\sigma^2\|\Delta^{\left( k
      \right)}\|_2^2/n<1$).



The lesson we learn from this motivating example is that
statistical estimation can benefit from leveraging
higher-order moment information, even when the moment
measurements are not essentially independent of each
other. This is particularly prominent for the random
corruption model, where all the ``higher-order trigonometric
moments'' in $H^{\left( k \right)}$, $k>1$ come from the
first moments in $H=H^{\left( 1 \right)}$ by taking
entrywise powers. In drastic contrast is the message-passing
algorithm in \cite{PWBM2018}, for which independence of the
complex Gaussian Wigner noise $\Delta^{\left( k \right)}$'s
across the frequency channels play an essential role. The
AMP approach was motivated by the \emph{non-unique games}
(NUG) framework in \cite{BCS2015}. Our algorithm follows an efficient
two-stage paradigm (\emph{initialization} and
\emph{iterative refinement}) popularized by recent progress
in nonconvex optimization (see, e.g. \cite{CLS2015,CC2015}),
and combines the trigonometric moments information across
frequency channels in a manner akin to classical
\emph{harmonic retrieval} techniques in signal processing
\cite{SM1997,TK1982,BM1986,ZW1988,Schmidt1986,RK1989,SDL2017a,SDL2017b}
and the \emph{generalized power method}
\cite{Boumal2016}. This strategy easily extends to
synchronization over general compact Lie groups, as
illustrated in Section~\ref{sec:extens-other-synchr}.

\vspace{-0.11in}
\paragraph{Notations} Upper case letters $A,B,C,\cdots$ and lower 
case letters $a,b,c,\cdots$ will be used to denote matrices
and vectors, respectively. $A^{*}$, $A^{\top}$ are the
transpose of $A$ with or without conjugation, respectively. 
The entrywise (Hadamard) product of matrix $A$ and $B$ will be denoted
as $A\circ B$. Graphs $G=\left(
  V,E \right)$ are always undirected and connected. Vertices
of the graph will be denoted as 
integers $1,2,\cdots,\left| V \right|$; pairs of integers $\left( i,j \right)$
denote edges in $E$.  For $n\in\mathbb{N}$ we write $\left[ n
\right]:=\left\{ 1,\cdots,n \right\}$.
Norms $\left\|
  \cdot \right\|_2$, $\left\| \cdot \right\|_{\infty}$
stand for matrix or vector norms, depending on the context;
$\left\| \cdot \right\|_{\mathrm{op}}$, $\left\| \cdot
\right\|_{\textrm{F}}$ are matrix operator and Frobenius
norms, respectively. The Cartesian product of $n$ copies of
$\Unitary \left( 1 \right)$ is denoted as
$\mathbb{C}_1^n$. The quotient space $\mathbb{R}/2\pi$ is
identified with the unit circle.


\section{Related Work}
\label{sec:related_work}

\paragraph{Phase synchronization}
Directly solving \eqref{eq:opt-prob-ang-sync} is
NP-hard \cite{ZH2006}, but many convex and nonconvex methods
have been proposed to find high quality approximate
solutions. These include spectral and semi-definite programming
(SDP) relaxations
\cite{Singer2011,CSC2012,CKS2015,BKS2016,BBS2017}. An
alternative approach using \emph{generalized power
  method} (GPM) is also studied
\cite{Boumal2016,LYM2017,ZB2018}.


\paragraph{Phase synchronization in multiple frequency channels}
\cite{BCS2015} proposed the \emph{non-unique games} (NUG)
SDP optimization framework for synchronization over compact Lie
groups.  The SDP is based on quadratically lifting the
irreducible representations of the group elements, and
imposing consistency among variables across frequency
channels via a F{\'e}jer kernel; it is computationally
expensive. \cite{PWBM2018} introduced an iterative
approximate message passing (AMP) algorithm for noise model
\eqref{eq:multi-freq-additive-gaussian}, assuming the noise
are Gaussian and independent across frequency channels. Each
iteration of the AMP performs matrix-vector multiplication
and entrywise nonlinear transformation, followed by an extra
Onsager correction term; it is conjectured to be
asymptotically optimal. \vspace{-0.1in}

\section{Algorithm}
\label{sec:algorithm}

In this section we formally state the two-stage
multi-frequency phase synchronization algorithmic
paradigm. Stage One combines phase synchronization outcomes from
individual frequency channels with harmonic retrieval, aiming
at producing a high-quality
initialization; Stage Two iteratively refines an input 
by an extended generalized power method that
works concurrently in multiple frequency channels while
striving to maintain entrywise consistency.




\subsection{Stage One: Initialization Strategy}
\label{sec:init-strat}

Our algorithm takes as input $k_{\mathrm{max}}$ Hermitian
measurement matrices $H^{\left( k \right)}$,
$k=1,\dots,k_{\mathrm{max}}$, arising from the general
sub-Gaussian model \eqref{eq:multi-freq-additive-gaussian}
(which includes \eqref{eq:multi-freq-random-corruption} as a
special case). This stage can be divided into three steps.



\noindent\textbf{Step~1.} \emph{Individual Frequency Synchronization}:
  Apply any phase synchronization algorithm (spectral/SDP
  relaxation or GPM) to get phase vector estimate $u^{\left( k
    \right)}\in\mathbb{C}^n$ from each $H^{\left( k \right)}$,
  $k=1,\cdots,k_{\mathrm{max}}$, and form $W^{\left( k \right)}\coloneqq
  u^{\left( k \right)}( u^{\left( k \right)} )^{*}$;

\noindent\textbf{Step~2.} \emph{Entrywise Harmonic
  Retrieval}: For each $\left( i,j \right)\in E$, use any
harmonic retrieval technique to estimate $\theta_i-\theta_j$
from $W^{\left( k \right)}_{ij}$,
$k=1,2,\cdots,k_{\mathrm{max}}$, call the estimators $\hat{\theta}_{ij}$;

\noindent\textbf{Step~3.} \emph{Final Phase
  Synchronization}: Construct another Hermitian matrix
$\widehat{H}\in\mathbb{C}^{n\times n}$ by
$\widehat{H}_{ij}\coloneqq e^{\iota \hat{\theta}_{ij}}$, and
apply any phase synchronization algorithm to estimate the
true phases $\{ e^{\iota \theta_1},\cdots,e^{\iota
    \theta_n} \}$ from matrix $\widehat{H}$.

\begin{algorithm}[bt]
   \caption{Periodogram Peak Extraction with Spectral
     Methods (PPE-SPC)}
   \label{alg:mf-angsync}
\begin{algorithmic}
   \STATE {\bfseries Input:} Hermitian matrices $\left\{ H^{\left( k
       \right)}\mid 1\leq k\leq k_{\mathrm{max}} \right\}$
   \STATE {\bfseries Output:} Initialization $\hat{x}\in \mathbb{C}_1^n$
   \STATE \underline{\textbf{Step 1: Individual Frequency Synchronization}}
   \FOR{$k=1$ {\bfseries to} $k_{\mathrm{max}}$}
   \STATE Extract the leading eigenvector $u^{\left(
       k\right)}$ of $H^{\left( k \right)}$ with scaling $\left\| u^{\left( k \right)} \right\|_2=\sqrt{n}$
    \STATE $W^{\left( k \right)}\gets u^{\left( k \right)}\left(u^{\left( k \right)}\right)^{*}$
   \ENDFOR
   \STATE \underline{\textbf{Step 2: Entrywise Harmonic Retrieval}}
   \FOR{$i=1$ {\bfseries to} $n$}
   \STATE
   $\displaystyle\hat{\theta}_{ij}\gets\argmax_{\phi\in
     \mathbb{R}/2\pi} \left|\mathrm{Re}\left\{
     \sum_{k=1}^{k_{\mathrm{max}}} W^{\left( k \right)}_{ij} e^{-\iota k\phi} \right\}\right|$
   \ENDFOR
   \STATE \underline{\textbf{Step 3: Final Phase Synchronization}}
   \STATE Construct Hermitian $\widehat{H}\in\mathbb{C}^{n\times n}$
   by $\widehat{H}_{ij}=e^{\iota
     \hat{\theta}_{ij}}$
   \STATE Extract the leading eigenvector $\hat{u}=\left(
     \hat{u}_1,\dots,\hat{u}_n \right)^{\top}$ of
   $\widehat{H}$
   \STATE $\hat{x}\gets ( \hat{u}_1/|\hat{u}_1|, \dots,  \hat{u}_n/|\hat{u}_n|) $
\end{algorithmic}
\end{algorithm}

The flexibility of the multi-frequency phase
synchronization framework lies at the various choices to be
made in each step. As a concrete example, we detail in
Algorithm~\ref{alg:mf-angsync}
a simple version that
uses spectral relaxation for phase synchronization and
periodogram-based harmonic retrieval. We will henceforth
refer to Algorithm~\ref{alg:mf-angsync} as the
\emph{periodogram peak extraction with spectral
methods} (PPE-SPC). If a different phase synchronization method
is used, for instance, SDP relaxation, our nomenclature
refers to it as PPE-SDP. We will focus on
analyzing PPE-SPC in depth in
Section~\ref{sec:analysis}, but the analysis strategy can be
seamlessly carried in principle to other variants of this
algorithmic paradigm. 


We briefly motivate the argmax operation in
Step~2  as follows. If our
measurement matrices are noise-free, then the $\left(  i,j
\right)$th entry of $W^{\left( k \right)}$ from Step~1
should equal to $e^{\iota k\left( \theta_i -
    \theta_j\right)}$; in this case, the goal of Step~2 is
to reconstruct $\left( \theta_i-\theta_j \right)$ from its
``trigonometric moments,'' for which any harmonic retrieval
technique can be applied; the periodogram method
in Algorithm~\ref{alg:mf-angsync} is among the most
na{\"i}ve approach for this purpose. For clean signal, the periodogram $|\mathrm{Re}\{\sum_{k=1}^{k_{\mathrm{max}}}
    W^{\left( k \right)}_{ij}e^{-\iota
      k\phi}\}|$ is equal to the modulus of the
Dirichlet kernel\vspace{-0.2in}
\begin{equation}
  \label{eq:dir-kernel}
  \begin{aligned}
    \mathrm{Dir}_{k_{\mathrm{max}}}\left( \theta_{ij}-\theta
  \right)&:=\sum_{k=-k_{\mathrm{max}}}^{k_{\mathrm{max}}}e^{\iota
  k \left( \theta_{ij}-\theta \right)}\\
         &=\frac{\sin \left[ \left(
          k_{\mathrm{max}}+1/2 \right) \left(
          \theta_{ij}-\theta \right)\right]}{\sin \left[ \left(\theta_{ij}-\theta\right) /2 \right]}
  \end{aligned}
\end{equation}
which attains its maximum at
$\theta_{ij}=\theta_i-\theta_j\left(\mathrm{mod}
  2\pi\right)$. Since the peak of
$\mathrm{Dir}_{k_{\mathrm{max}}}$ becomes sharper and
sharper as $k_{\mathrm{max}}$ increases, we expect the
periodogram peak identification step to be robust to noise,
which will produce a very high quality estimate
$\widehat{H}$ for Step~3. In fact, our analysis in
Section~\ref{sec:analysis} suggests that this initialization
stage alone can produce highly accurate phase vectors for
sufficiently large $k_{\mathrm{max}}$, and the estimation
error drops inverse-polynomially in $k_{\mathrm{max}}$.

\subsection{Stage Two: Iterative Refinement}
\label{sec:iterative-algorithms}
In this stage, we use an iterative refinement scheme that takes an initial phase vector and enhances it successively. In our implementation we warm-start this iterative algorithm with the $\hat{x}$ produced from the PPE-SPC Algorithm~\ref{alg:mf-angsync}, but any initialization scheme can be applied in principle, including random initialization. This iterative refinement concurrently performs the generalized power method (GPM) \cite{Boumal2016} in multiple frequency channels consistently: at each frequency $k$, we perform power iteration by multiplication with $H^{\left( k \right)}$; the results are combined across frequency channels to obtain one periodogram for each vertex $i$ followed by a ``soft harmonic retrieval'' step that soft-thresholds \cite{donoho1995noising} the periodogram in frequency domain. We pick a relatively lower threshold at the beginning of this iterative scheme, but gradually raise the threshold over $0.99$ to reveal the true peak that persists. Details can be found in Algorithm~\ref{alg:mf-gpm}, henceforth referred to as \emph{multi-frequency generalized power method} (MFGPM).







\begin{algorithm}
   \caption{Multi-Frequency Generalized Power Method (MFGPM)}
   \label{alg:mf-gpm}
\begin{algorithmic}
   \STATE {\bfseries Input:} Hermitian matrices $\left\{ H^{\left( k
       \right)}\mid 1\leq k\leq k_{\mathrm{max}} \right\}$, initialization $\hat{x}\in\mathbb{C}_1^n$; threshold $\tau$
   \STATE {\bfseries Output:} Phase vector $\hat{z}\in\mathbb{C}_1^n$
      \FOR{$k=1$ {\bfseries to} $k_{\mathrm{max}}$}
      \STATE $z^{\left( k,0 \right)}\gets \hat{x}^k$
   \ENDFOR 
   \FOR{ $t = 1$ {\bfseries to} $T$}
   \FOR{$k=1$ {\bfseries to} $k_{\mathrm{max}}$}
   \STATE $u^{(k,t)}\gets H^{(k)} z^{(k,t-1)}$. 
    \ENDFOR 
   \FOR{$i=1$ {\bfseries to} $n$}
   \STATE $\displaystyle h^{(t)}_i(\theta) = \mathrm{Re} \left\{\sum_{k  = 1}^{k_\m} u^{(k,t)}_{i} e^{\imath k \theta} \right\}$
   \STATE Soft-thresholding: $\hat{u}_i^{(k,t)} \!\!\gets\!\! \int_0^{2\pi}\!\!\!\! \eta_\tau(h^{(t)}_i(\theta)) e^{-\imath k \theta} \mathrm{d}\theta $ 
    \STATE Rounding: $z_i^{(k,t)} \gets \hat{u}_i^{(k,t)}\big/|\hat{u}_i^{(k,t)}|$ 
         \ENDFOR 
    \ENDFOR 
    \STATE $\hat{z}\gets z^{(1,T)}$
\end{algorithmic}
\end{algorithm} 

MFGPM can be viewed as an iterative version of PPE-SPC, except that the stringent peak extraction step is replaced with the more malleable soft-thresholding. Periodograms $h_i^{\left( t \right)}$ are virtually the Dirichlet kernels, which truncate a Dirac delta function in the frequency domain; one can also take Ces{\'a}ro means of these periodograms, or equivalently, work with the \emph{F{\'e}jer kernels} that are known to converge faster to the Dirac delta function. We omit those results as no significant difference is observed in performance.

\begin{figure}
\begin{tabular}{c}
\includegraphics[width =0.49\columnwidth]{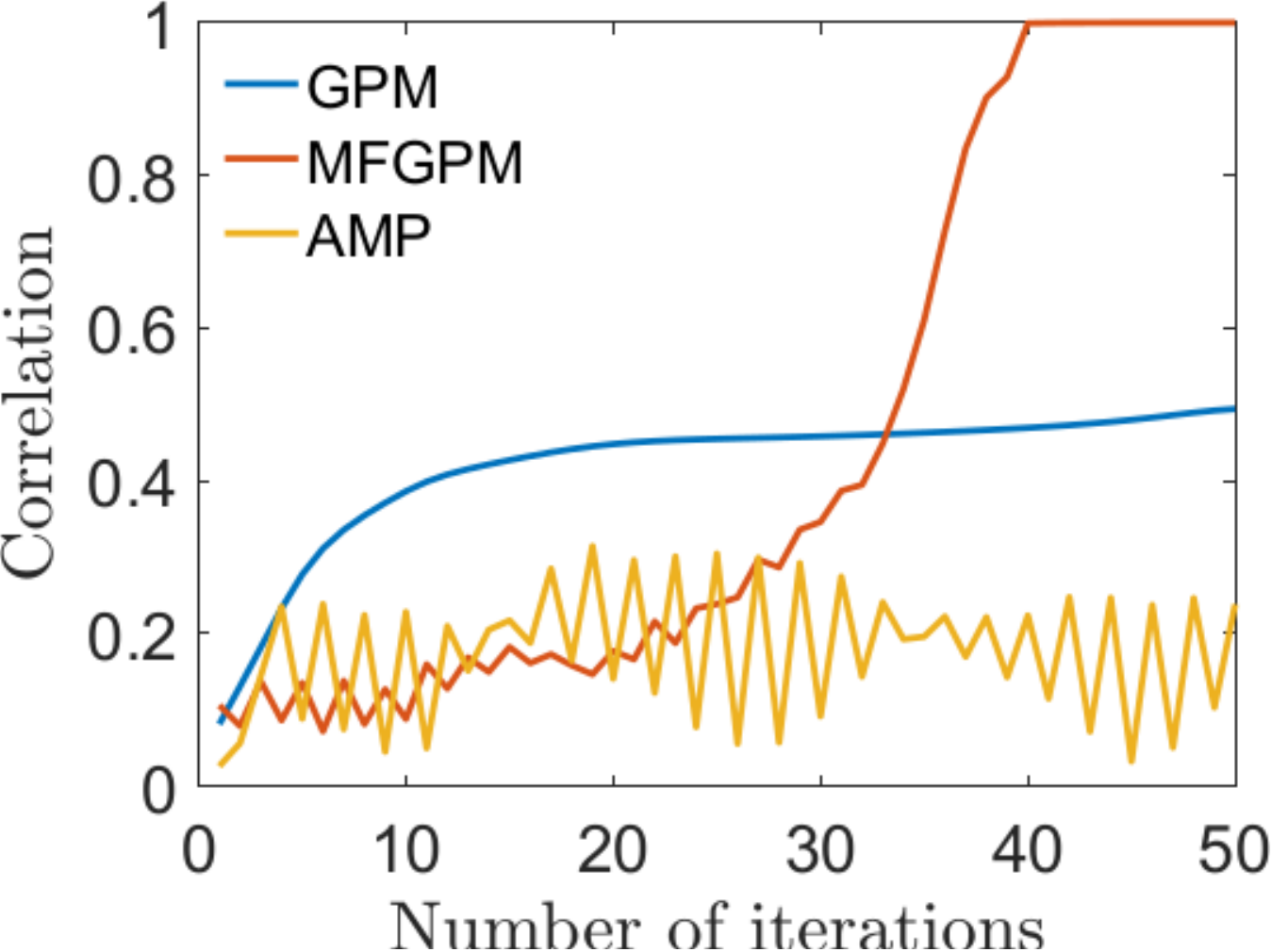}
\includegraphics[width =0.49\columnwidth]{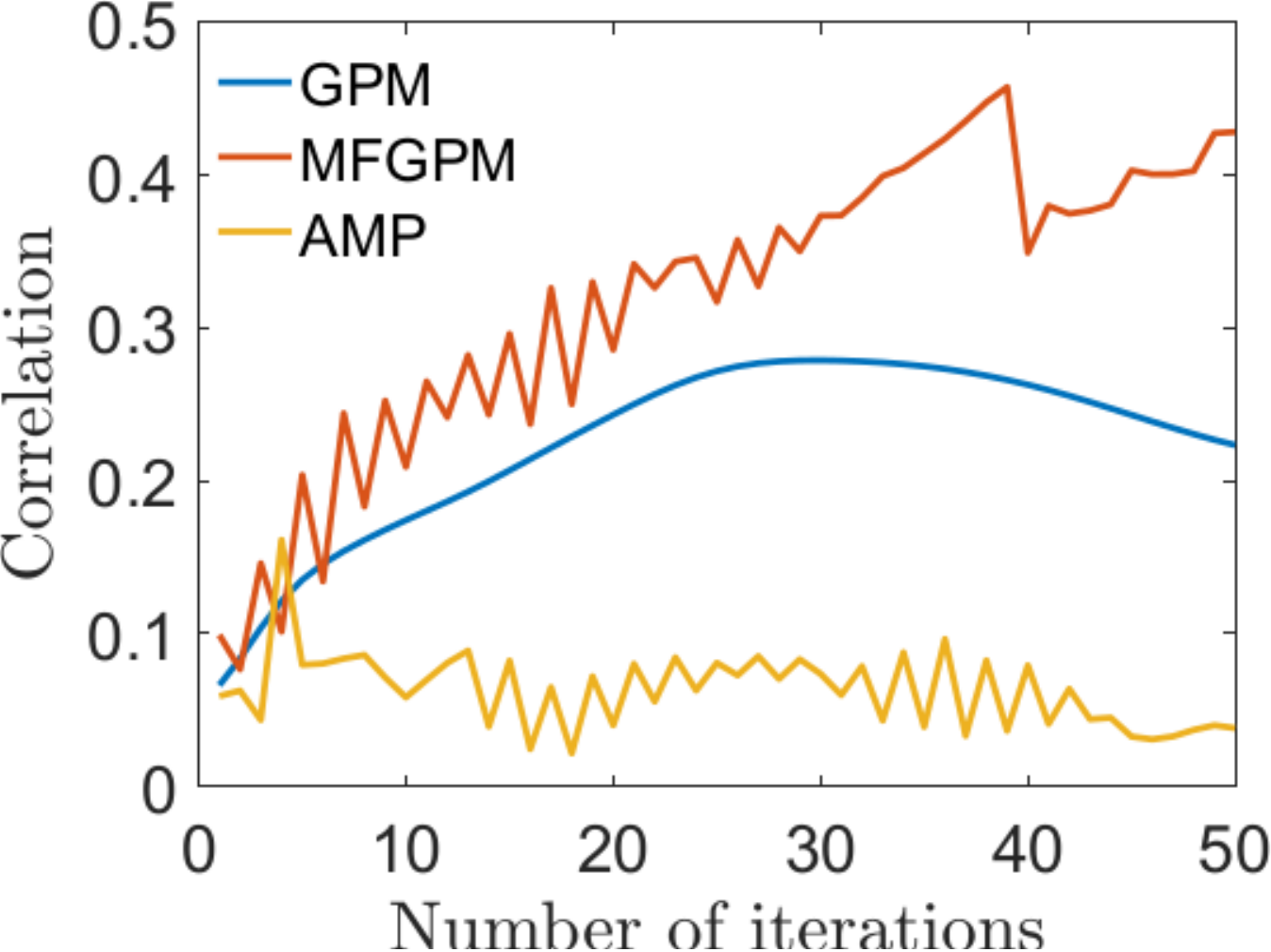}
\end{tabular}
\caption{Evolution of correlations between iterates and the true phase vector $z\in\mathbb{C}_1^n$ for three iterative algorithms with the same random initialization. The three iterative algorithms in comparison are generalized power method (GPM, \cite{Boumal2016}), multi-frequency generalized power method (MFGPM, Algorithm~\ref{alg:mf-gpm}), and approximate message passing (AMP, \cite{PWBM2018}). The figure illustrates a typical success run (\textbf{left}) where MFGPM started with a random initialization of low correlation (around $0.1$) but terminated with an output of correlation $>0.5$, along with a failure run (\textbf{right}) started with another random initialization of comparable correlation but terminated with correlation $<0.5$. In both circumstances, outputs from MFGPM attain higher correlation than GPM or AMP. Input data are generated from the random corruption model \eqref{eq:multi-freq-random-corruption} with $r = 0.1$ and $n = 100$. See Figure~\ref{fig:n100_cg_rc} for more systematic comparison results under this noise model.}
\label{fig:iter}
\end{figure}

As an integral part of our two-stage algorithmic framework, MFGPM works most efficiently with initialization from PPE-SPC, but we also observed empirically that the MFGPM outperforms other methods given identical random initialization, illustrated in Figure~\ref{fig:iter}, in the sense that MFGPM often produces phase vectors that correlate more strongly with the true phase vector $z$. See Section~\ref{sec:numer-exper} for more comprehensive comparisons results.




The computational complexities of PPE-SPC and MFGPM are $\mathcal{O}\left( k_{\m}n^3 \right)$ and $\mathcal{O}\left( Tk_{\m}n^2 \right)$, respectively.

\section{Analysis}
\label{sec:analysis}

In this section we analyze PPE-SPC in theory, under the
general sub-Gaussian noise model
\eqref{eq:multi-freq-additive-gaussian}. We assume the
observation graph $G$ is generated from a
Erd\H{o}s--R{\'e}nyi model with edge connectivity $p\in
\left[ 0,1 \right]$ independent of the $\Delta^{\left( k \right)}$'s.


\begin{assumption}
\label{assum:noise-sub-gaussian}
For $\sigma>0$ and each $k\in [k_{\mathrm{max}}]$, assume
\begin{equation}
  \label{eq:randcor-multifreq}
  H^{\left( k \right)}=A\circ [z^k \left( z^k \right)^{*}+\sigma\Delta^{\left( k \right)}]
\end{equation}
where $z^k\in\mathbb{C}_1^n$ is the entrywise $k$th
power of $z$, and $\Delta^{\left( k \right)}$, $k=1,\dots,k_{\mathrm{max}}$
are complex random Wigner matrices satisfying the following assumptions:
\begin{enumerate}[(1)]
\item\label{item:1} For any fixed $k\in [k_{\m}]$, $\{\mathrm{Re}(\Delta^{\left( k \right)}_{\ell
        j}), \mathrm{Im}(\Delta^{\left( k \right)}_{\ell
        j})\mid 1\leq \ell <j\leq n\}$ are jointly
      independent with zero
    mean, and unit sub-Gaussian norm \cite{Vershynin2018}; 
\item\label{item:2} $\Delta_{ii}^{\left( k \right)}=0$ for
  all $1\leq k\leq k_{\mathrm{max}}$ and $1\leq i\leq n$;
\item\label{item:3} $\Delta^{\left( k \right)}_{\ell
    j}=\overline{\Delta^{\left( k \right)}_{j\ell}}$ for all $k=1,\dots,k_{\mathrm{max}}$ and $\ell<j$.
\end{enumerate}
Furthermore, assume $A$ is the adjacency matrix of a
Erd\H{o}s--R{\'e}nyi random graph independent of
all the $\Delta^{\left( k \right)}$'s, with edge connecting
probability $p\in \left[ 0,1 \right]$.
\end{assumption}
We emphasize again that
Assumption~\ref{assum:noise-sub-gaussian} assumes no 
independence for the $\Delta^{\left( k \right)}$'s across
frequency channels; only entries within the same $\Delta^{\left( k
  \right)}$ are assumed independent. As explained in
Introduction, this enables us to unify our discussions on
the random corruption model and additive Gaussian model in a
single pass (see e.g.,
\eqref{eq:random-corruption-example}). Another advantage for
such generality is that we can focus on analyzing complete observation graphs, since\vspace{-0.05in}
\begin{equation*}
  \mathbb{E}H^{\left( k \right)}=pz^k \left( z^k \right)^{*}-pI_n\vspace{-0.05in}
\end{equation*}
where $I_n$ is the identify matrix of dimension
$n$-by-$n$, and thus we can apply the theoretical analysis
in this section to $\frac{1}{p}\left( H^{\left( k \right)}+pI_n \right)=z^k\left(z^k\right)^{*}+E^{\left( k \right)}$
where\vspace{-0.05in}
\begin{equation*}
  E^{\left( k \right)}=\frac{1}{p}\left\{ A\circ [z^k \left(
      z^k \right)^{*}+\sigma\Delta^{\left( k
      \right)}]\right\}-z^k \left( z^k \right)^{*}+I_n\vspace{-0.05in}
\end{equation*}
satisfies the same conditions as $\Delta^{\left( k \right)}$ in
Assumption~\ref{assum:noise-sub-gaussian} with different
absolute constants. Therefore, in the rest of this section
we focus on complete observation graph $G$ only, i.e.,
\begin{equation}
\label{eq:complete-graph-noise-model}
  H^{\left( k \right)}=z^k \left( z^k
  \right)^{*}+\sigma\Delta^{\left( k \right)}, \quad
  1\leq k\leq  k_{\mathrm{max}}.\vspace{-0.1in}
\end{equation}

Our first goal is to understand the spectral method in
PPE-SPC Step~1 and Step~3. Since
Step~2 is entrywise, it is crucial to bound the
$\ell_{\infty}$ distance between $z^k$ and the leading eigenvector
$u^{\left( k \right)}$ (scaled to $\| u^{\left( k \right)}
\|_2=\sqrt{n}$). The
proof of the following Lemma~\ref{lem:basic-ell-infty-bound}
uses recent $\ell_{\infty}$ perturbation results of eigenvectors of
random matrices \cite{EBW2017,AFWZ2017,FWZ2018,ZB2018}
and can be found in the supplemental material.

\begin{lemma}
\label{lem:basic-ell-infty-bound}
  Assume Assumption~\ref{assum:noise-sub-gaussian} is
  satisfied, and the observation graph $G$ is a complete
  graph. Let $\epsilon\in \left( 0,2 \right]$ be an
  arbitrarily chosen but fixed absolute constant. For any
  $k\in [ k_{\mathrm{max}} ]$, denote
  $u^{\left( k \right)}$ for the leading eigenvector of
  $H^{\left( k \right)}$ scaled such that $\left\| u^{\left(
        k \right)} \right\|_2=\sqrt{n}$ and $(
  z^k)^{*}u^{\left( k \right)}=|( z^k )^{*}u^{\left( k
    \right)}|$. There exist absolute (in particular, independent of $k$ and
  $n$) constants $c_0, C_0,C_2>0$ such that, if
  $\sigma<c_0\sqrt{n/\log n}$, there holds with probability
  $1-\mathcal{O}\left( n^{-\left( 2+\epsilon \right)} \right)$
  \begin{align}
    \label{eq:lemma-ell-infty}
      \| u^{\left( k \right)}-z^k \|_{\infty}&\leq
    C_0\sigma\sqrt{\log n / n},\\
      \label{eq:recon-offset-bound}
   \left| W_{ij}^{\left( k \right)}-z_i^k\bar{z}_j^k \right|&\leq
   C_2\sigma\sqrt{\log n / n}.
  \end{align}
\end{lemma}
The inequality \eqref{eq:recon-offset-bound} is a direct
consequence of \eqref{eq:lemma-ell-infty}, which is
identical to Theorem~8 of \cite{ZB2018}, but we verify in
the proof that the event probability $1-\mathcal{O}(
  n^{-2} )$ in \cite{ZB2018} can be made slightly
higher. This is necessary for
taking the union bound across all $\mathcal{O}( n^2 )$ entries in the main Theorem~\ref{thm:main-thm}.

A quick consequence of Lemma~\ref{lem:basic-ell-infty-bound}
is the uniform proximity of the periodogram to a Dirichlet
kernel up to constant scaling and shifts, with high probability. More
specifically,
\begin{align*}
    &\Bigg| \mathrm{Re}\left\{\sum_{k=1}^{k_{\mathrm{max}}} W^{\left( k \right)}_{ij}
    e^{-\iota
  k\phi}\right\}-\frac{1}{2}\left[\mathrm{Dir}_{k_{\mathrm{max}}}\left(
                  \theta_i-\theta_j-\phi\right)-1\right]\Bigg|\notag\\
    &\leq \left|\sum_{k=1}^{k_{\mathrm{max}}}
      \left(W^{\left( k \right)}_{ij}- z_i^k\bar{z}^k_j\right)
    e^{-\iota k\phi}\right| \leq 2C_2
      k_{\mathrm{max}}\sigma\sqrt{\log n/n}
  \end{align*}
 with probability $1-\mathcal{O}\left(
  n^{-\left( 2+\epsilon \right)} \right)$. Clearly, the
maximum of
$\left|\mathrm{Dir}_{k_{\mathrm{max}}}\left(\theta_i-\theta_j-\phi\right)-1\right|$
is attained at $\theta=\theta_i-\theta_j$.
We thus expect the argmax operation in Step~2 of
PPE-SPC to produce high accuracy
estimates of $\theta_i-\theta_j$ as long as the difference
between the ``optimization landscape'' of the periodogram
and the Dirichlet kernel is small enough. This is formalized
in the following lemma, which exploits the geometry of the
Dirichlet kernel.

\begin{lemma}
  \label{lem:peak-loc-estimates}
Under the same conditions as in
Lemma~\ref{lem:basic-ell-infty-bound}, if
\begin{equation}
  \label{eq:gap-control-condition}
    \left[2k_{\mathrm{max}}\sin \left(
        \pi/\left(2k_{\mathrm{max}}+1\right) \right)
    \right]^{-1}+4C_2\sigma\sqrt{\log n / n}<1
\end{equation}
then with probability
at least $1-\mathcal{O}\left( n^{-\left( 2+\epsilon \right)}
\right)$
\begin{equation}
  \label{eq:peak-loc-estimates}
  \left| \hat{\theta}_{ij}-\left( \theta_i-\theta_j \right)
  \right|\leq 4\pi/\left(2k_{\mathrm{max}}+1\right).
\end{equation}
\end{lemma}

It is straightforward to check that
\eqref{eq:gap-control-condition} holds for
sufficiently large $k_{\mathrm{max}}$ as long as
$4C_2\sigma\sqrt{\log n/n}$ is bounded from above by
$1-1/\pi$. This can be seen by noticing that the function $\left[2x\sin \left( \pi/ \left(
      2x+1 \right) \right)\right]^{-1}$ is differentiable and monotonically
decreasing for all $x\geq 2$, and for sufficiently large
$k_{\mathrm{max}}$ it infinitesimally approaches $1/\pi<1$.

The most important message from
Lemma~\ref{lem:peak-loc-estimates} is the following: At the
beginning of the Step~3 of
PPE-SPC, the newly
constructed Hermitian matrix $\widehat{H}$ is entrywise
$\mathcal{O} \left( k_{\mathrm{max}}^{-1} \right)$--close to
the ground truth rank-one matrix $zz^{*}$. We emphasize
that this error incurred in $\widehat{H}$
is significantly smaller than the noise level $\sigma$ in the raw
input data, and can be made arbitrarily small by choosing large
$k_{\mathrm{max}}$. We
formalize this key observation in the main
theorem below, for which the proof is deferred to the
supplemental material.

\begin{theorem}
\label{thm:main-thm}
  Under the same conditions as
  Lemma~\ref{lem:basic-ell-infty-bound} and
  Lemma~\ref{lem:peak-loc-estimates}, if
  \eqref{eq:gap-control-condition} holds and $4c_0C_2<1-\sqrt{2}/\pi$,
  then there exists an
  absolute constant $C_3>0$ such that, with probability
  $1-\mathcal{O}\left( n^{-\epsilon} \right)$, the
  correlation between the true phase vector $z$ and the
  leading eigenvector $\hat{u}$ (scaled to $\left\| \hat{u} \right\|_2=\sqrt{n}$) of $\widehat{H}$ in
  PPE-SPC Step~3 is at least
  \begin{equation}
    \label{eq:key-error-estimate}
      \mathrm{Corr} \left( \hat{u}, z \right)\geq 1-C_3/k_{\mathrm{max}}^2\vspace{-0.1in}
  \end{equation}
provided that
\begin{equation*}
  k_{\mathrm{max}} > \max\left\{5, \left(\sqrt{2}\,\pi \left( 1-4 C_2\sigma\sqrt{\log n / n}\right)-2\right)^{-1}\right\}.
\end{equation*}
Moreover, for the phase vector
 $\hat{x}$ output
  from PPE-SPC,
  \begin{equation*}
    \mathrm{Corr}\left( \hat{x}, z \right)\geq 1-4C_3/k_{\mathrm{max}}^2.
  \end{equation*}
\end{theorem}

Following the discussion after
Lemma~\ref{lem:peak-loc-estimates}, it is not surprising to
see in Theorem~\ref{thm:main-thm} that the correlation can
be made arbitrarily close to $1$ (or equivalently, the
$\ell_2$ distance between the estimated and true phase
vectors can be made arbitrarily close to $0$). Moreover, it doesn't take
excessively large $k_{\mathrm{max}}$ for
PPE-SPC to outperform all existing
phase synchronization algorithms---in fact, for
$\sigma\asymp\mathcal{O}( \sqrt{n/\log n} )$ which is the
highest level of noise tolerable to ensure the validity of
Lemma~\ref{lem:basic-ell-infty-bound}, it suffices to take
$k_{\mathrm{max}}=\mathcal{O}\left( \sqrt{n}/\sigma
\right)\asymp\mathcal{O}\left( \sqrt{\log n} \right)$
to suppress the $\ell_2$ estimation error below the
established near-optimal bound $\mathcal{O}\left( \sigma
\right)$ for eigenvector based phase synchronization
methods \cite{BBS2017,ZB2018}. We believe
\eqref{eq:key-error-estimate} can still be improved by a
factor of $\sqrt{n}$ by leveraging the randomness in the
residue error in \eqref{eq:recon-offset-bound}, but such
finer analysis relies on more detailed analysis on the
$\ell_{\infty}$ perturbation and the change in the
optimization landscape, which will be pursued in a
future work.

\section{Extension to General Synchronization}
\label{sec:extens-other-synchr}

The algorithmic framework of multi-frequency phase synchronization proposed in this paper can be extended to synchronization over any compact Lie group $\mathcal{G}$, by the representation-theoretic analogue of Fourier series --- the \emph{Peter--Weyl decomposition}. In a nutshell, the \emph{Peter--Weyl theorem} states that, for square integrable functions $f\in L^2 \left( \mathcal{G} \right)$, we have decomposition
\begin{equation}
\label{eq:IFTG}
f(g) = \sum_{k=0}^{\infty} d_k \mathrm{tr} \left( \hat{f}(k) \rho_k(g) \right)
\end{equation}
where each $\rho_k:\mathcal{G}\rightarrow\mathbb{C}^{d_k\times d_k}$ is an irreducible, unitary representation of $\mathcal{G}$, and $\hat{f}\left( k \right)$ is the ``Fourier coefficient''
\begin{align}
\label{eq:groupFT}
\hat{f}(k) = \int_{\mathcal{G}} f(g)\rho_k(g)\,\mathrm{d}g,
\end{align}
where the integral is take with respect to the Haar measure.

On a connected observation graph $G$, the input data to a \emph{synchronization problem over group $\mathcal{G}$} are pairwise measurements $g_{ij}\in\mathcal{G}$ on edges $\left( i,j \right)\in E$ satisfying $g_{ij}=g_{ji}^{-1}$. The goal is to find $n$ group elements $g_1,\dots,g_n\in\mathcal{G}$, one for each vertex, that satisfy as many constraints $g_{ij}=g_ig_j^{-1}$ as possible. Mathematically, this type of problems can often be formulated as an optimization problem \cite{BCS2015}\vspace{-0.1in}
\begin{equation}
\label{eq:NUG}
\min_{g_1,\dots,g_n\in\mathcal{G}}\sum_{i, j = 1}^n f_{ij}(g_i g_j^{-1}), \vspace{-0.05in}
\end{equation} 
where each $f_{ij}\in L^2 \left( \mathcal{G} \right)$ measures the compatibility between the relative alignment $g_i g_j^{-1}$ and the observation data $g_{ij}$ on edge $\left( i,j \right)\in E$. The $f_{ij}$'s are nonlinear and nonconvex in general. If $f_{ij}$ are bandlimited, we can expand \eqref{eq:NUG} using the Peter--Weyl decomposition
\begin{align*}
    \sum_{i, j = 1}^nf_{ij}(g_i g_j^{-1})= \sum_{k=0}^{k_\m}\sum_{i, j = 1}^n  d_k \mathrm{tr}\left[ \hat{f}_{ij}(k) \rho_k(g_i) \rho^*_k(g_j) \right]
\end{align*}
which can be viewed as a generalization of the multi-frequency phase synchronization problem \eqref{eq:opt-prob-ang-sync-mult-freq}.

For simplicity of statement, we assume the observation graph $G$ is complete in this section. Since $\rho_k$'s are unitary representations, the matrices $\rho_k \left( g \right)$'s are unitary matrices for any $g\in \mathcal{G}$, and it is natural to solve for $g_i$ from its irreducible representations $\rho_k \left( g_i \right)$. Vertically stacking the $k$th irreducible representations together, the variable can be organized in matrices $X^{\left( k \right)}\in \mathbb{C}^{nd_k\times d_k}$, $k\in\mathbb{Z}$ defined by
\begin{equation}
  \label{eq:xkgroup}
  X^{(k)} = \left[\rho_k(g_1), \dots,\rho_k(g_n) \right]^{\top}.
\end{equation}
Analogies of the noise models also exist in this more general setting. The additive Gaussian noise model, following \cite{PWBM2018}, amounts to\vspace{-0.1in}
\begin{equation}
\label{eq:noise_gau_rho}
H^{(k)} = \frac{\lambda_{k}}{n} X^{(k)} (X^{(k)})^* + \frac{1}{\sqrt{n d_k} } \Delta^{(k)}\vspace{-0.1in}
\end{equation}
where the parameter $\lambda_k>0$ stands for the signal-to-noise ratio (SNR) at ``frequency $k$,'' $\Delta^{(k)}\in\mathbb{C}^{nd_k\times nd_k}$ is a Wigner matrix with i.i.d. standard complex Gaussian entries in the upper triangular part. For the random corruption model, let
\begin{equation}
\label{eq:g_rc_model}
g_{ij} = \begin{cases}
g_i g_j^{-1}, & \text{with probability $r$}\\
\tilde{g}\sim\mathrm{Unif}\left( \mathcal{G} \right), &  \text{with probability $1-r$}
\end{cases}
\end{equation}
and set the $(i, j)$th sub-block of $H^{(k)}$ to $\rho_k (g_{ij})$.

As we elaborate in the remainder of this section, all the key ingredients in PPE-SPC and MFGPM can be extended to this more general setting. We demonstrate the efficacy of this algorithm for $\SO \left( 3 \right)$ synchronization in Section~\ref{sec:numer-exper}.

\noindent\textbf{Spectral relaxation: } Compute the top $d_k$ eigenvectors and stack them horizontally to form $U^{(k)} = [ u^{(k)}_1, \dots, u^{(k)}_{d_{k}} ]$. Approximate $H^{(k)}$ with $\widehat{H}^{(k)} = U^{(k)} \left(U^{(k)}\right)^*$. 


\noindent \textbf{Generalized harmonic retrieval: } 
For each $\left( i,j \right)\in E$, set\vspace{-0.1in}
\begin{equation}
\label{eq:best_g}
\hat{g}_{ij} = \argmax_{g \in \mathcal{G}}\sum_{k = 1}^{k_\m}  d_k \mathrm{tr}\left[ \widehat{H}_{ij}^{(k)} \rho^*_k(g) \right]. \vspace{-0.1in}
\end{equation}
Based on these new estimates for the pairwise alignments, we build matrix $\widehat{H}$ with $n^2$ blocks with $\widetilde{H}_{i j} = \rho_1(\hat{g}_{ij})$. We then extract the top $d_1$ eigenvectors of $\widehat{H}$, stack them horizontally to form $\widetilde{U} = [ u_1, u_2, \dots u_{d_1} ]$, and project each of its $n$ vertical blocks $\widetilde{U}_1\dots, \widetilde{U}_n\in\mathbb{C}^{d_1\times d_1}$ to a unitary matrix through singular value decomposition (SVD)
\begin{equation}
\label{eq:rounding}
\mathrm{Proj}(\widetilde{U}_i ) = \Phi \Psi^*\quad  \text{where} \quad \widetilde{U}_i = \Phi \Sigma \Psi^*.
\end{equation}



\noindent \textbf{Iterative refinement: } 
At the $t$th iteration, denoting $X^{(k,t)}$ for the current stacked $k$th representations \eqref{eq:xkgroup}, we construct
\begin{equation*}
Y^{(k)} = H^{(k)} X^{(k,t)},
\end{equation*}
and compute the inverse Fourier transform for each of the $n$ vertical sub-blocks $Y_1^{(k)},\dots, Y_n^{(k)}\in\mathbb{C}^{d_k\times d_k}$ of $Y^{(k)}$: 
\begin{equation*}
h_i(g) = \sum_{k = 1}^{k_\m} \mathrm{Re}\left\{ d_k \mathrm{tr}\left[ Y_i^{(k)} \rho^*_k(g) \right] \right\},\quad i=1,\dots,n.
\end{equation*}
Note that we only need toe evaluate $C_i(g)$ on a finite number of uniformly sampled elements of $\mathcal{G}$, from which the ``inverse Fourier transform'' can be applied 
\begin{equation}
\label{eq:soft_th_g}
U_i^{(k)} =  \int_{\mathcal{G}} \eta_\tau\left(h_i(g)\right) \rho_k(g)\,\mathrm{d} g
\end{equation}
along with the soft-thresholding $\eta_{\tau}$. We again project each $U_i^{(k)}$ to the closest unitary matrix by SVD \eqref{eq:rounding}, then form $X^{(k,t+1)}$ by vertically stacking the $\mathrm{Proj}( U_i^{\left( k \right)} )$'s. The final outputs are $\widehat{X}^{\left( k \right)}=X^{\left( k,T \right)}$ for $k=1,\dots,k_{\m}$.




  


\begin{figure}[t!]
\captionsetup[subfloat]{farskip=2pt,captionskip=1pt}
\begin{center}
\subfloat[AMP]{
\includegraphics[width = 0.33\columnwidth]{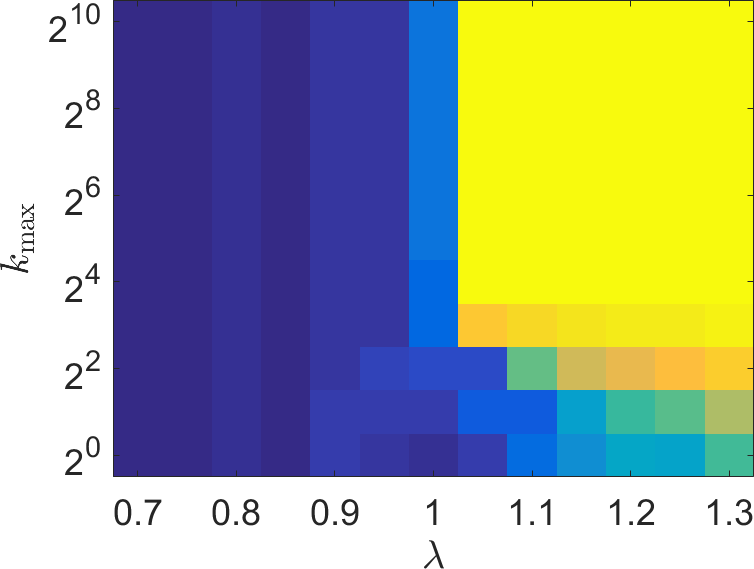}
\label{fig:gau_cg_AMP}
}
\subfloat[PPE-SPC]{
\includegraphics[ width = 0.33\columnwidth]{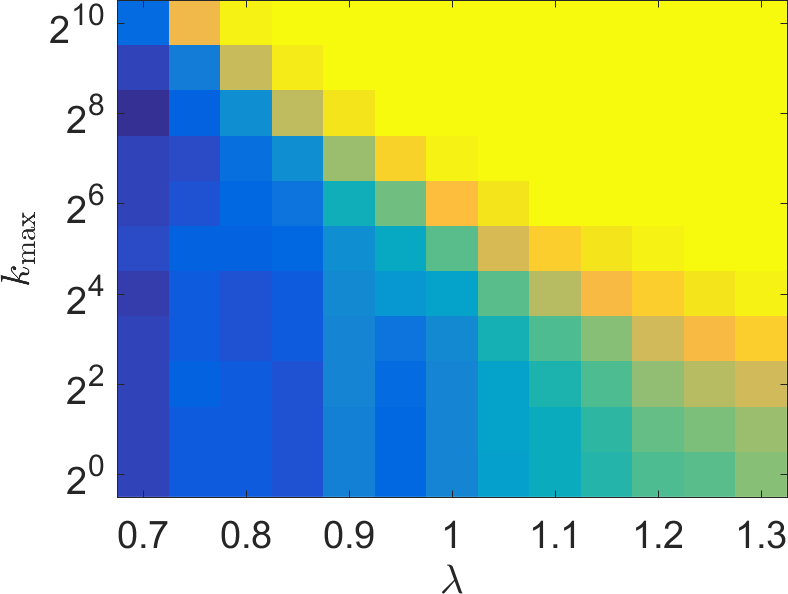}
\label{fig:gau_cg_spec1}
}
\subfloat[ MFGPM ]{
\includegraphics[ width = 0.32\columnwidth]{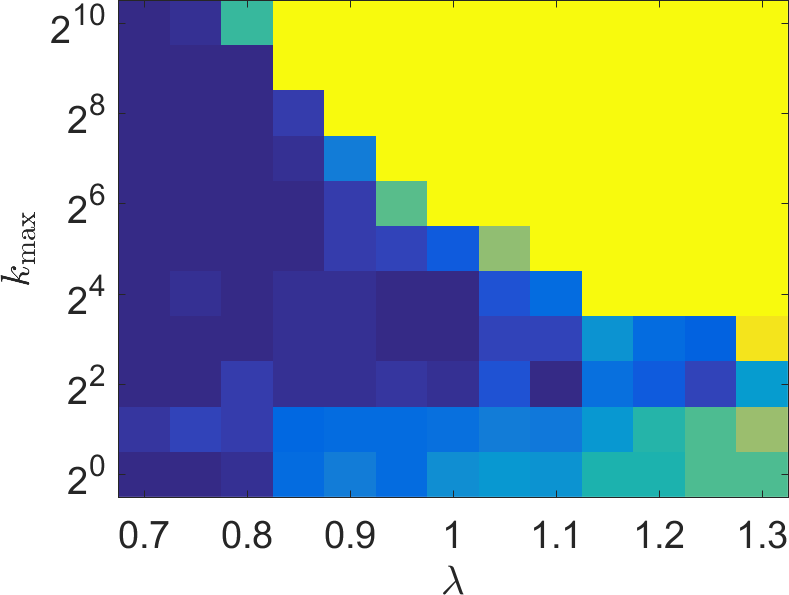}
\label{fig:gau_cg_projpower}
}\\
\subfloat[PPE-SDP]{
\includegraphics[width = 0.32\columnwidth]{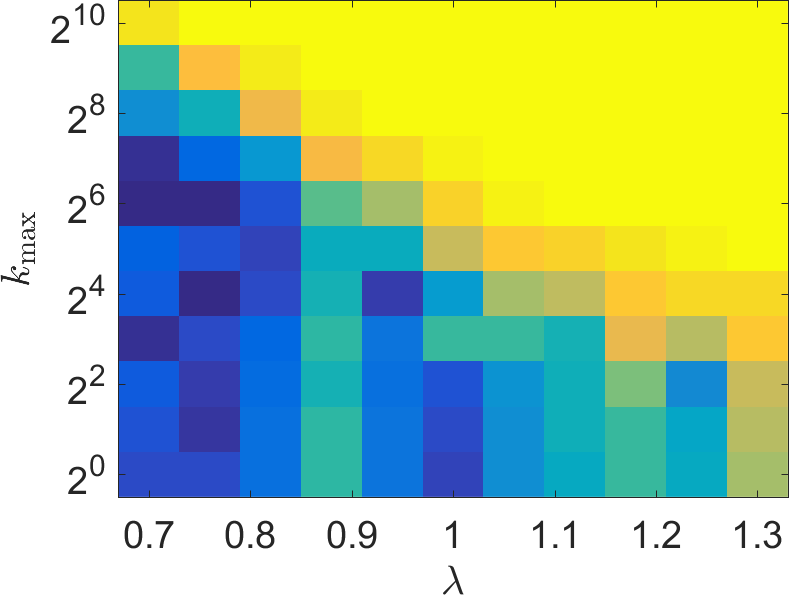}
\label{fig:gau_cg_sdp1}
}
\subfloat[PPE-SDP]{
\includegraphics[ width = 0.32\columnwidth]{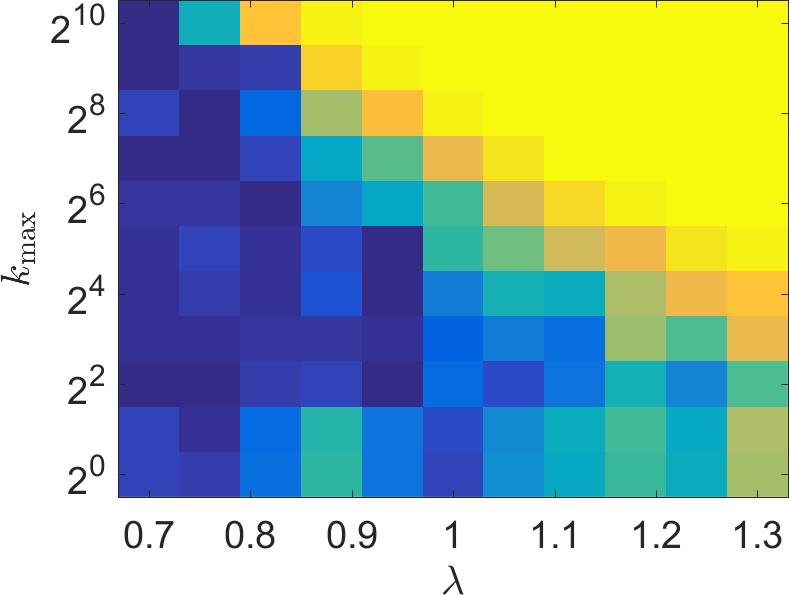}
\label{fig:gau_cg_sdp}
} 
\subfloat[PPE-SPC$^3$]{
\includegraphics[ width = 0.32\columnwidth]{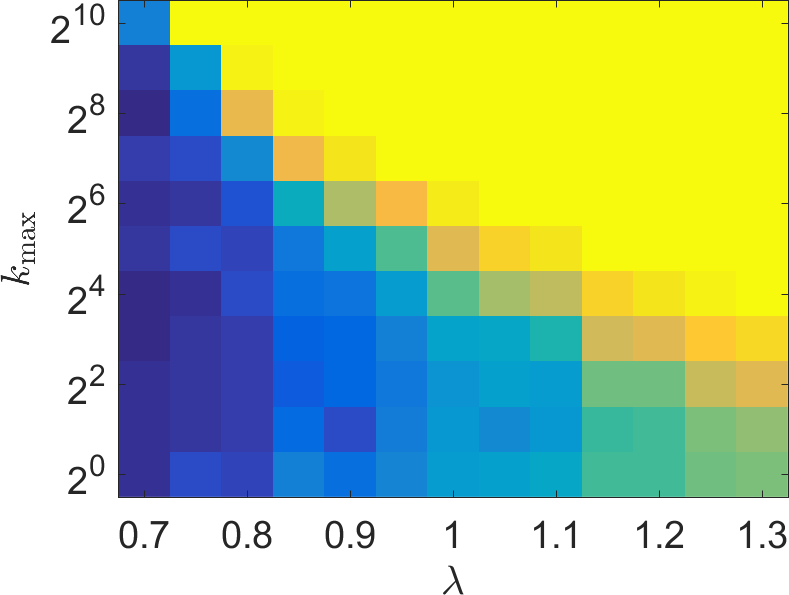}
\label{fig:gau_cg_spec2}
}\\
\flushleft
\subfloat[PPE-SPC + AMP]{
\includegraphics[ width = 0.32\columnwidth]{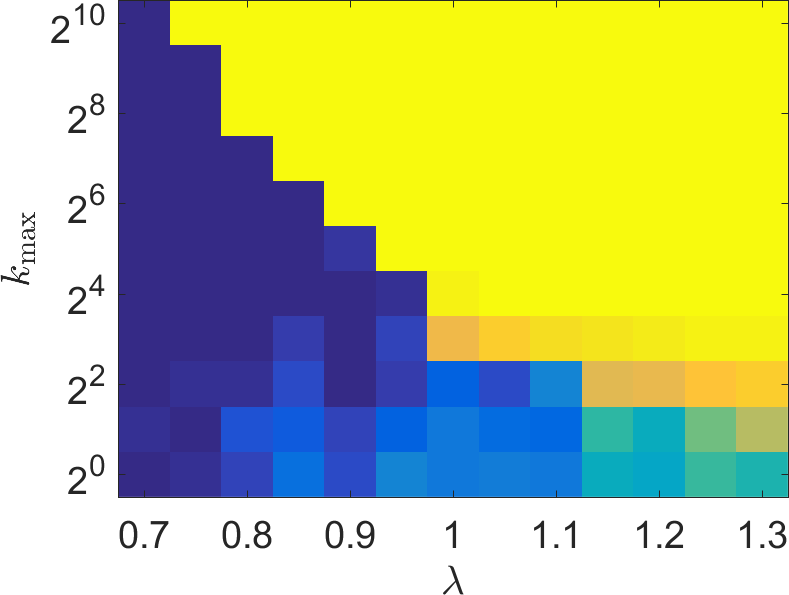}
\label{fig:gau_cg_AMP_winit}
} 
\subfloat[ PPE-SPC + MFGPM ]{
\includegraphics[width = 0.32\columnwidth]{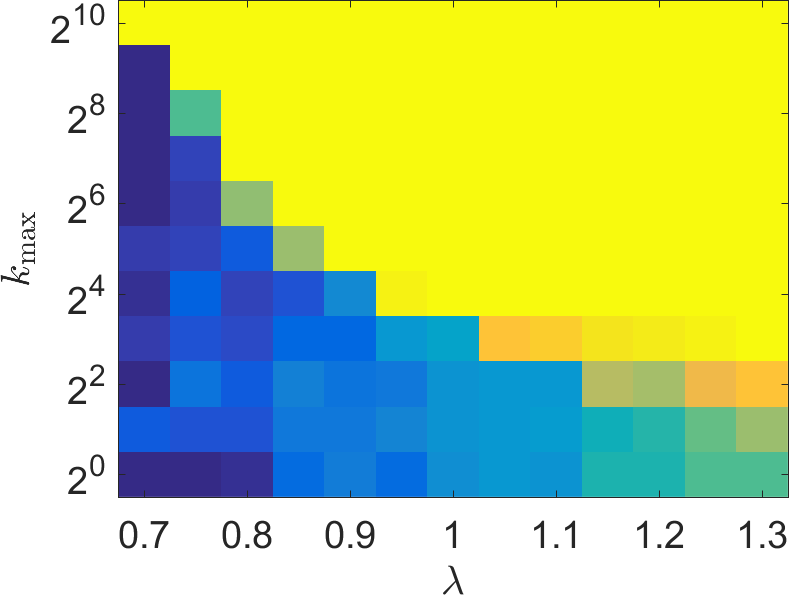}
\label{fig:gau_cg_projpower_winit}
}  \quad \quad \quad 
\includegraphics[height=0.25\columnwidth]{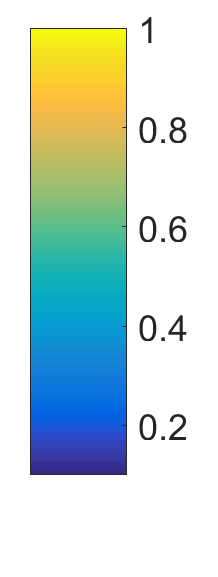}
\end{center}
\caption{$\Unitary\left(1\right)$ synchronization under Gaussian noise model for $n=100$. Here $\sigma=\sqrt{n}/\lambda$. Every data point is a median over 40 trials in  \protect \subref{fig:gau_cg_AMP}--\protect \subref{fig:gau_cg_projpower} and \protect \subref{fig:gau_cg_spec2}--\protect \subref{fig:gau_cg_projpower_winit},  and over 5 trials in \protect \subref{fig:gau_cg_sdp1}--\protect \subref{fig:gau_cg_sdp}. }
\label{fig:n100_cg_gau}
\end{figure}

\section{Numerical Experiments}
\label{sec:numer-exper}
This section contains detailed numerical results under both additive Gaussian noise and random corruption models, for both $\Unitary \left( 1 \right)$ and $\SO \left( 3 \right)$. In all experiments with Gaussian noise, we keep $\sigma_k\equiv \sigma\equiv \sqrt{n}/\lambda$ where $\lambda>0$ is the signal-to-noise ratio (SNR); for the random corruption model \eqref{eq:random-corruption-phase} we set $r \equiv \lambda/\sqrt{n}$. We fix $n=100$ and vary $\lambda$ and $k_\m$ to evaluate and compare the performance of different algorithms. When comparing iterative algorithms (AMP, GPM, MFGPM), within each random trial the random initialization is kept identical for all three algorithms and across frequency channels; between trials both data and initialization are redrawn. The remainder of the section contains results for $\Unitary \left( 1 \right)$ and $\SO \left( 3 \right)$ synchronization with complete observation graphs only; incomplete observation graph results are similar and included in the supplemental material.

\begin{figure}[t!]
\captionsetup[subfloat]{farskip=2pt,captionskip=1pt}
\begin{center}
\subfloat[AMP]{
\includegraphics[ width = 0.32\columnwidth]{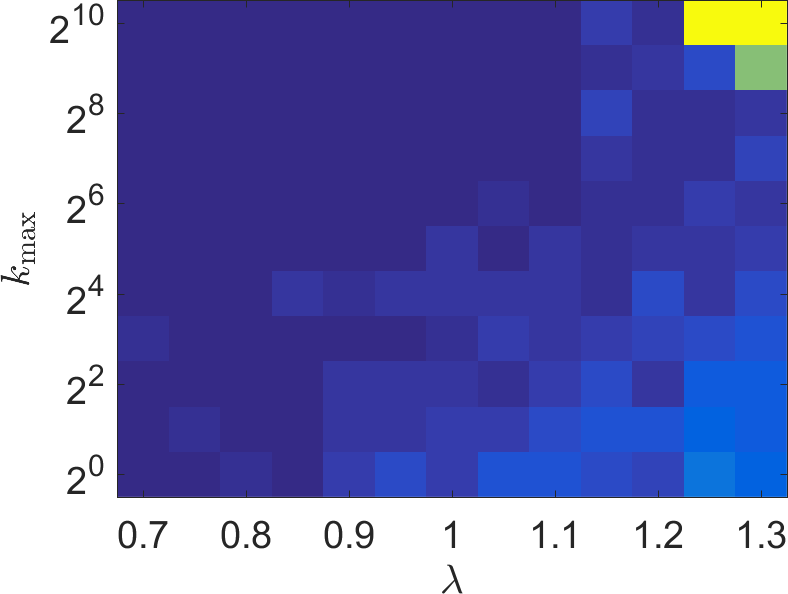}
\label{fig:rc_cg_AMP}
}
\subfloat[PPE-SPC]{
\includegraphics[ width = 0.32\columnwidth]{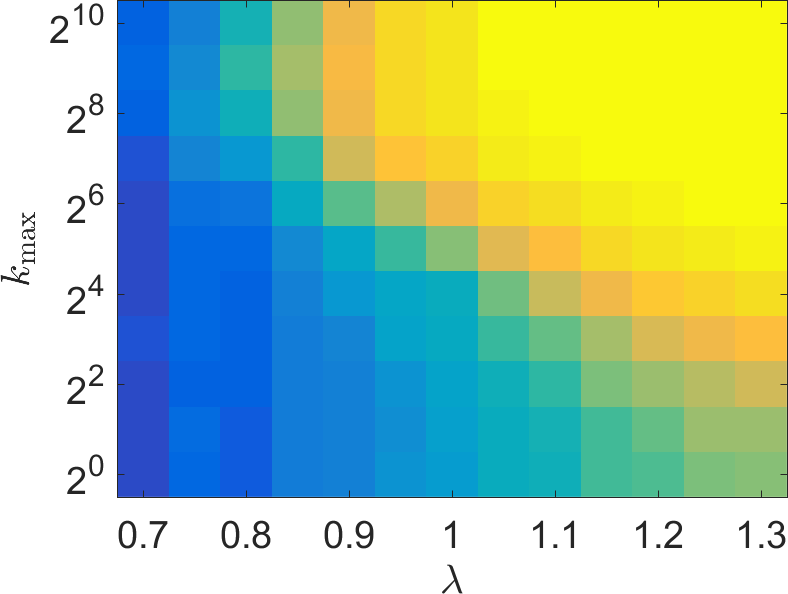}
\label{fig:rc_cg_spec1}
}
\subfloat[MFGPM]{
\includegraphics[ width = 0.32\columnwidth]{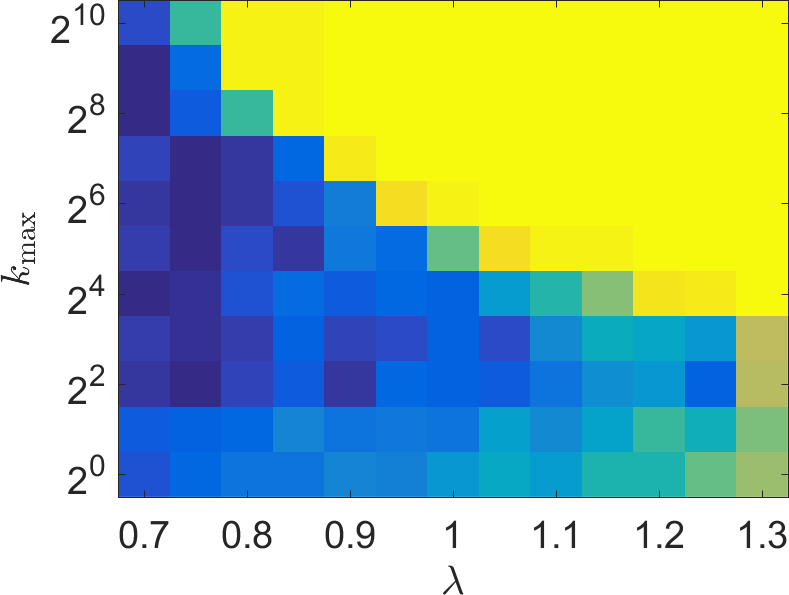}
\label{fig:rc_cg_projpower}
} \\
\subfloat[PPE-SDP]{
\includegraphics[ width = 0.32\columnwidth]{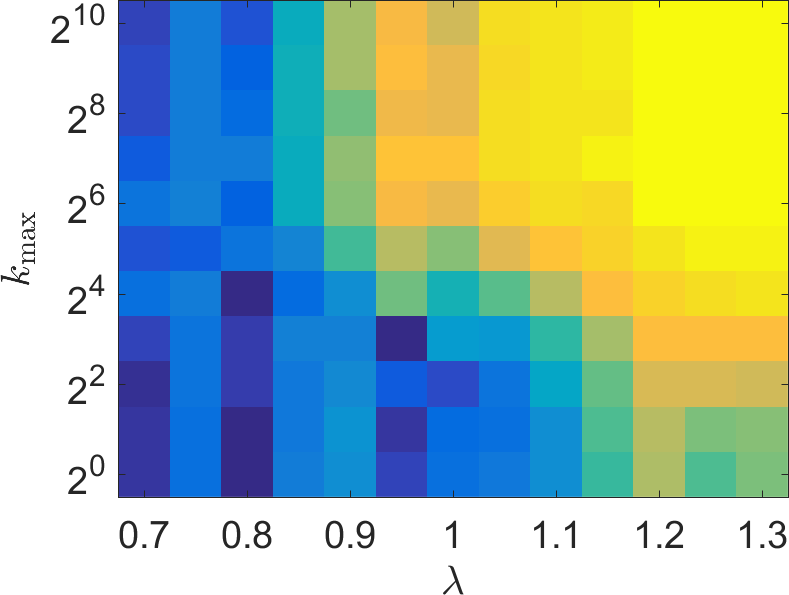}
\label{fig:rc_cg_sdp1}
} 
\subfloat[PPE-SDP]{
\includegraphics[ width = 0.32\columnwidth]{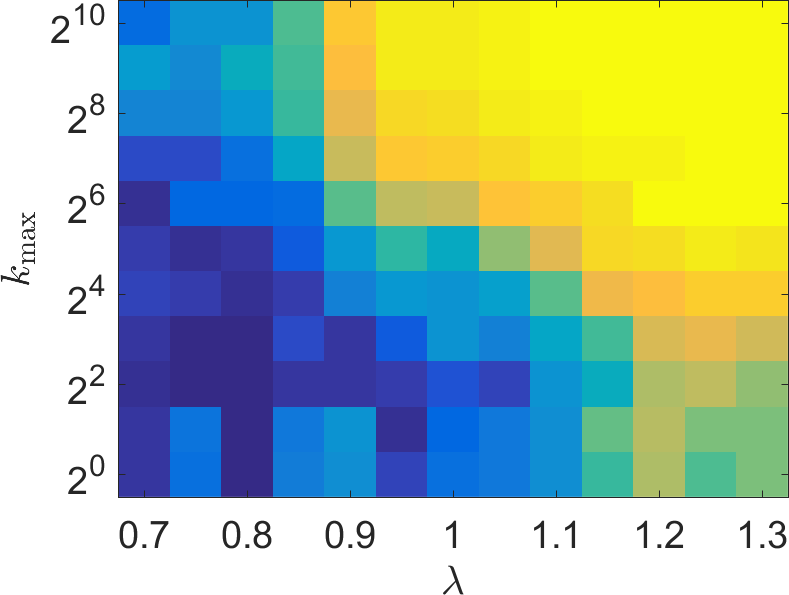}
\label{fig:rc_cg_sdp}
}
\subfloat[PPE-SPC$^3$]{
\includegraphics[ width = 0.32\columnwidth]{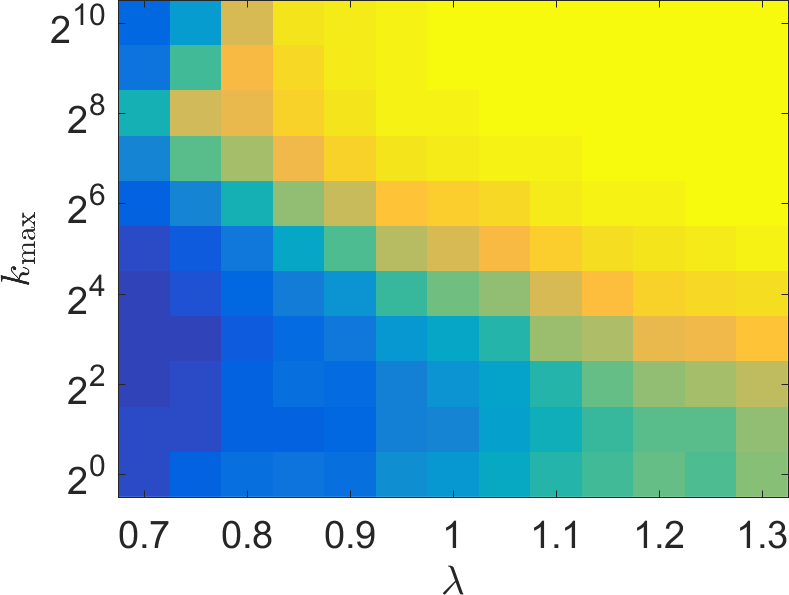}
\label{fig:rc_cg_spec2}
} \\
\flushleft 
\subfloat[PPE-SPC + AMP]{
\includegraphics[width = 0.32\columnwidth]{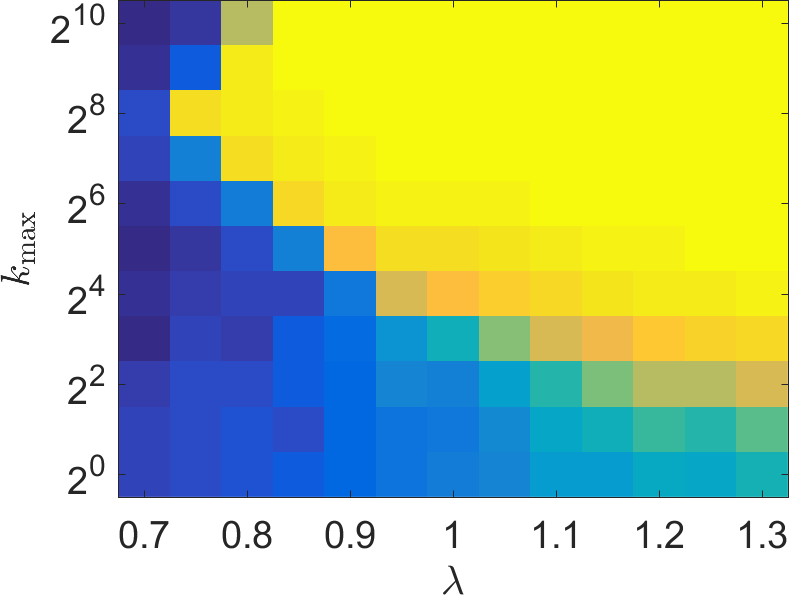}
\label{fig:rc_cg_AMP_winit}
} 
\subfloat[PPE-SPC + MFGPM]{
\includegraphics[width = 0.32\columnwidth]{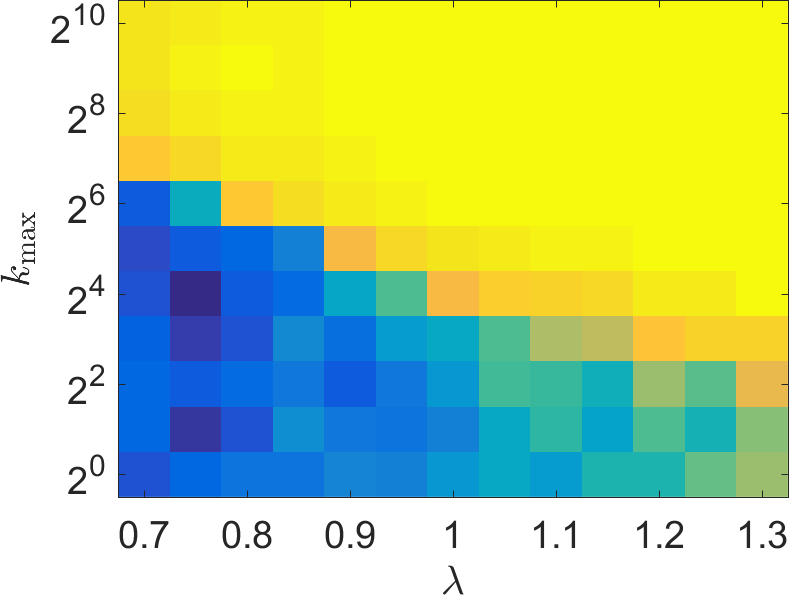}
\label{fig:rc_cg_projpower_winit}
}\quad \quad \quad
\includegraphics[height=0.25\columnwidth]{./figures/color_bar}
\end{center}
\caption{$\Unitary\left(1\right)$ synchronization under random corruption model for $n = 100$. Here $r = \lambda/\sqrt{n}$. Every data point is a median over 40 trials in  \protect \subref{fig:rc_cg_AMP}--\protect \subref{fig:rc_cg_projpower} and \protect \subref{fig:rc_cg_spec2}--\protect \subref{fig:rc_cg_projpower_winit},  and over 5 trials in \protect \subref{fig:rc_cg_sdp1}--\protect \subref{fig:rc_cg_sdp}.} 
\label{fig:n100_cg_rc}
\end{figure}

\noindent \textbf{$\Unitary \left( 1 \right)$ synchronization: }
In Figure~\ref{fig:n100_cg_gau} and Figure~\ref{fig:n100_cg_rc}, we measure the correlation between the output and the truth phase vector for various single- and multi-frequency synchronization methods, under the additive Gaussian and random corruption noise model, respectively. The SNR $\lambda$ varies between $0.7$ and $1.3$, which is in the extremely noisy regime: under the random corruption model, for instance, with $n=100$,  between $87\% $ and $93\%$ of the pairwise alignments are corrupted with random elements. In each subplot, the vertical axis varies $k_{\mathrm{max}}$ from $1$ to $1024$, and the horizontal axis marks the change in $\lambda$. The bottom row in each subplot thus represents the single-frequency ($k_{\mathrm{max}}=1$) version of the algorithm. The methods under comparison are: (a) AMP \cite{PWBM2018} with random initialization; (b) PPE-SPC; (c) MFGPM with random initialization; (d) PPE-SDP (replacing the spectral methods in Algorithm~\ref{alg:mf-angsync} with SDP relaxation); (e) PPE-SDP with an additional projection to rank-one matrices in each iteration; (f) Iterating PPE-SPC three times; (g) AMP initialized with PPE-SPC; (h) MFGPM initialized with PPE-SPC.


It is clear from Figure~\ref{fig:n100_cg_gau} and Figure~\ref{fig:n100_cg_rc} that leveraging information in multiple frequency channels produces superior results than single-frequency approaches. Most shockingly, in Figure~\ref{fig:n100_cg_rc} our proposed PPE-SPC method and variants [subplots (b)--(h)] are capable of recovering the true phase vector when the SNR is well below the critical threshold $\lambda=1$ (corresponding to $r<1/\sqrt{n}$) determined in \cite{Singer2011} by random matrix arguments. This is surprising because, according to \cite{Singer2011}, for single frequency phase synchronization one can not expect correlation to be much higher than $1/\sqrt{n}$, which is $0.1$ in our experiments. This is confirmed by looking at the bottom row of each subplot of Figure~\ref{fig:n100_cg_rc}, but with suitably large $k_{\m}$ this barrier no longer exists, even though in model \eqref{eq:multi-freq-random-corruption} our high-frequency measurements are generated from the single frequency data.

\begin{figure}
\captionsetup[subfloat]{farskip=2pt,captionskip=1pt}
\begin{center}
\subfloat[Gaussian Noise Model]{
\includegraphics[width=0.48\columnwidth]{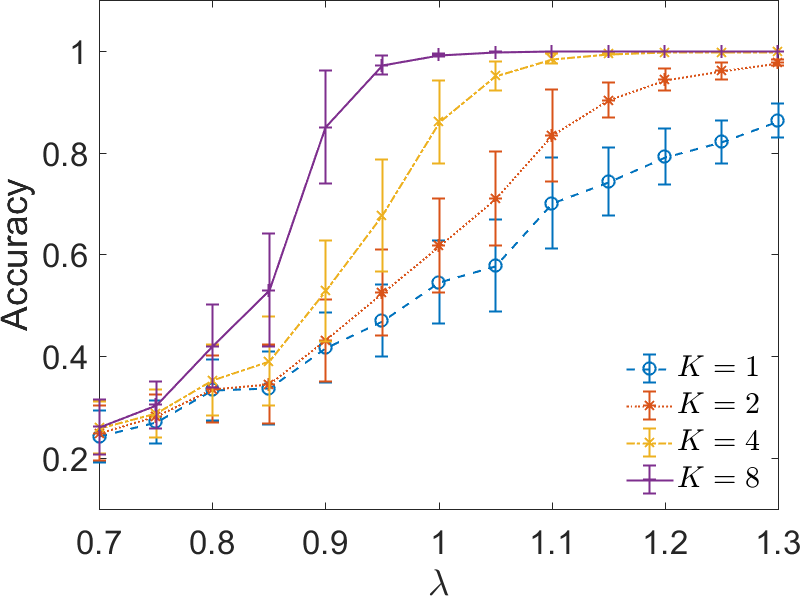}
\label{fig:SO3g}}
\subfloat[Random Corruption Model]{
\includegraphics[width=0.48\columnwidth]{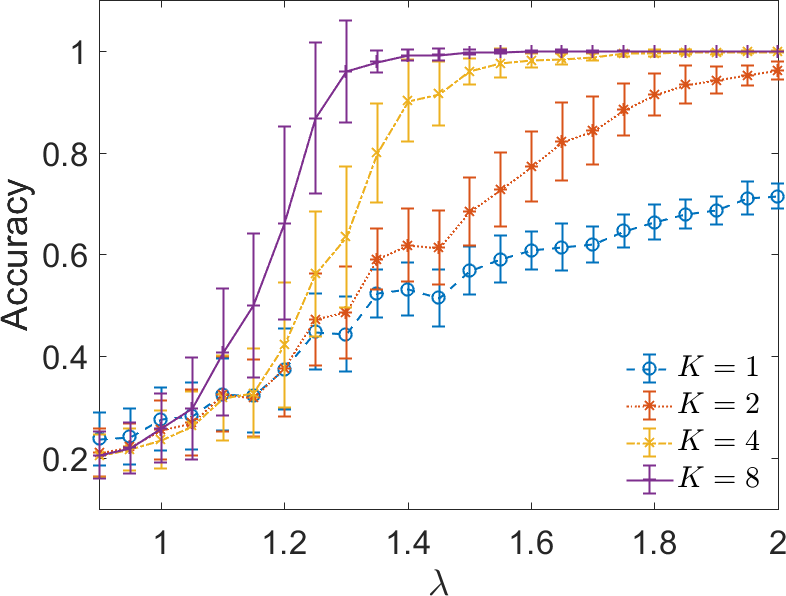}
\label{fig:SO3rc}}
\end{center}
\caption{Accuracy and error-bars for $\SO\left(3\right)$ synchronization with Gaussian noise (\textbf{left}) and random corruption (\textbf{right}) model with $K$ frequencies, for various $K$. The noise levels are kept as $\lambda = \sqrt{n} / \sigma $ for Gaussian models and $r=\lambda/\sqrt{n}$ for random corruption models, where $n = 100$. Accuracy is measured by the correlation $\| (X^{(1)} )^* \widehat{X}^{(1)} \|_F/(\sqrt{3} n)$ between estimates and the ground truth. Each data point is the median of $50$ trials. \vspace{-0.15in}}
\label{fig:SO3init}
\end{figure}

In Figure~\ref{fig:n100_cg_gau} and Figure~\ref{fig:n100_cg_rc}, (d) and (e) illustrates the performance of the SDP variant of PPE-SPC. The difference between (d) and (e) is the following: in (d) we use directly estimated $W^{(k)}$  by solving the SDP in \cite{Singer2011}, but in (e) we apply project the SDP solution to a rank-one matrix using eigen-decomposition. The results from these SDP variants are occasionally slightly better PPE-SPC, but the computational cost is expensive: the runtime is over $40$ times longer, and a lot more memory is required. The SDP relaxation in \cite{BCS2015} is even more demanding on computation resources so is not included here.

Figures~\ref{fig:gau_cg_spec2} and~\ref{fig:rc_cg_spec2} explore another possibility of extending PPE-SPC: After recovering $\widehat{H}$, take entrywise powers of $\hat{H}$ and treat them as multi-frequency data input to another fresh run of PPE-SPC. Unlike the iterative refinement algorithm MFGPM, we observed empirically that the performance boost saturate quickly after just a couple of such repeated calls to PPE-SPC. The result in (f) from both figures are obtained from performing $3$ such repetitions. Compared with (b), this strategy improves the estimation accuracy for smaller $\lambda$, but the performance gain is not as significant as using MFGPM for iterative refinements (h).




Initialization turns out to be important for AMP: As shown in Figure~\ref{fig:gau_cg_AMP}, when the SNR is below the critical threshold predicted in \cite{PWBM2018} ($\lambda<1$), increasing $k_{\mathrm{max}}$ does not lead to performance improvement; the critical threshold appears even higher for random corruption model (Figure~\ref{fig:rc_cg_AMP}). In contrast, PPE-SPC and MFGPM can always benefit from sufficiently larger $k_{\m}$.
 

\noindent \textbf{$\SO\left(3\right)$ synchronization: } Comparison results for $\SO \left( 3 \right)$ synchronization under Gaussian noise model and random corruption model are shown in Figure~\ref{fig:SO3g} and \ref{fig:SO3rc}, respectively. In all these experiments, the Fourier transform \eqref{eq:soft_th_g} is numerically evaluated using $m = 1000$ elements uniformly sampled in $\SO \left( 3 \right)$. Clearly, the proposed method outperforms single frequency methods and achieve higher accuracy as $k_{\m}$ increases; moreover, the multi-frequency formulation and algorithm lead to drastic performance boost especially at the  ``low SNR regime.''

\begin{figure}
\captionsetup[subfloat]{farskip=2pt,captionskip=1pt}
\begin{center}
\includegraphics[width=0.9\columnwidth]{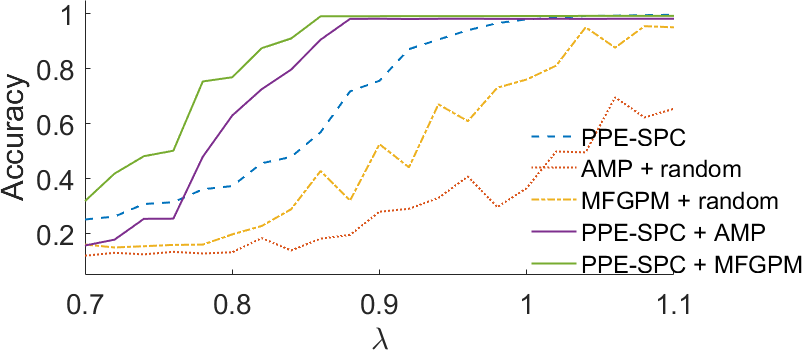}
\end{center}
\vspace{-0.5cm}
\caption{Comparison of random and spectral initialization for $\SO\left(3\right)$ synchronization, under noise model \eqref{eq:noise_gau_rho} where $k_\m = 8,n=100$. Each data point is averaged from $20$ random trials.}
\label{fig:SO3iter}
\end{figure}

In Figure~\ref{fig:SO3iter} we compare AMP and MFGPM with different initialization strategies--PPE-SPC vs. random initialization--under the additive Gaussian noise model \eqref{eq:noise_gau_rho} with $k_{\m}=8$. We plot the accuracy of using PPE-SPC alone without iterative refinement as a baseline. The results demonstrate the performance boost from using PPE-SPC for initialization, as well as improvements gain from using iterative refinements on top of the initialization PPE-SPC.






\section{Conclusion}
\label{sec:concl}

In this paper, we propose a novel, mult-frequency
formulation for phase synchronization as a nonconvex
optimization problem, for which we develop a two-stage
algorithm inspired by harmonic retrieval and generalized
power method that produces high accuracy
approximate solutions. We demonstrate in theory and
experiments that the new framework
significantly outperform all existing phase synchronization
algorithms.

There are many opportunities for future research. We are
particularly interested in gaining deeper theoretical
understandings for the multi-frequency GPM algorithm,
especially its performance guarantees and behavior near
local optimum. More general harmonic retrieval techniques
can be potentially used in place of the periodogram-based
peak extraction. We are also working on extending the
algorithmic framework beyond compact Lie groups, such as
Euclidean groups and symmetric groups, with applications to
object matching \cite{NIPS2016_6128,pachauri2013solving}.


\section*{Acknowledgements}
\label{sec:ack}

Tingran Gao acknowledges
support from an AMS-Simons Travel Grant and partial support
from DARPA D15AP00109 and NSF IIS 1546413.


\appendix

\section{Technical Proofs}
\label{sec:proofs}

\subsection{Proof of Lemma~1}
\label{sec:proof-lemma-1}

\begin{proof}[Proof of Lemma~1]
  The conclusion of this lemma is
identical to Theorem~8 of \cite{ZB2018}; the only difference
is that the event probability is slightly larger --- in
Theorem~8 of \cite{ZB2018} the event probability is
$1-\mathcal{O}\left( n^{-2} \right)$. This can be done by
straightforwardly modifying the arguments in the proof
of the Theorem~8 of \cite{ZB2018}, and at the expense of increasing the
absolute constant picked in that proof. Actually, this is
already stated by the authors of \cite{ZB2018} on page 998
of the published version, in the paragraph right below
their Theorem 5. We document here how this modification can
be done.

The randomness in the proof of Theorem~8 of \cite{ZB2018}
arises only from the dependence of Lemma~9 and Lemma~10 of
\cite{ZB2018}, so it is sufficient to track the failure
probability of the events there. These modifications only
need to be stated for real sub-Gaussian random variables, as
the trivial passage from real to complex cases is the same
as detailed in the proof of Lemma~9 of \cite{ZB2018}.

Lemma~9 of \cite{ZB2018} is based on the well-known
concentration results on the maximum singular value of
sub-Gaussian random matrices, in particular, 
Proposition~2.4 of \cite{RV2010}, which states for any sub-Gaussian random
matrix $A$ of dimension $n$-by-$n$ with independent, zero
mean sub-Gaussian entries (whose subgaussian moments are
bounded by $1$) that, for any $t>0$,
\begin{equation*}
  \mathbb{P}\left( \sigma_{\mathrm{max}}\left( A
    \right)>C\sqrt{n} +t \right) \leq 2e^{-ct^2}
\end{equation*}
where $c,C>0$ are positive absolute constants. We take here
$t=C\sqrt{n}$, so $\left\| A \right\|_2\lesssim\sqrt{n}$
with probability at least $1-2e^{-cC^2 n}$. Obviously, there
exists sufficiently large absolute constant $C_2>0$ such
that
\begin{equation*}
  e^{-cC^2n}\leq \frac{C_2}{n^{2+\epsilon}}\quad\forall n\in\mathbb{N},
\end{equation*}
where $\epsilon\in \left( 0,2 \right]$ is the arbitrarily chosen but fixed
constant in the statement of our Lemma~1.

Lemma~10 of \cite{ZB2018} attains the event probability
$1-\mathcal{O}\left( n^{-2} \right)$ by taking a union bound,
over $n$ instances of $1\leq m\leq n$ and $\left|
  \mathcal{U}_m \right|$ instances of $u\in\mathcal{U}_m$,
for individual event probabilities of $1-4e
n^{-5}-4e^{-c_2n/4}$, where $c_2$ is an absolute positive
constant. However, note that in the case of eigenvectors, we
have $\left| \mathcal{U}_m \right|=1$ (consisting of a
singleton, cf.~the second paragraph on pp.1000 of
\cite{ZB2018}, right above section title ``Introducing
auxiliary eigenvector problems''), which is two orders of
magnitude smaller than the bound $\left| \mathcal{U}_m
\right|\leq 3n^2$ stated in Lemma~10 of \cite{ZB2018}. The
union bound thus yields the success probability of at least
$1-4en^{-4}-4ne^{-c_2n/4}$, which is $1-\mathcal{O}\left(
  n^{-4} \right)$.

Combining both ends lead to the success probability of
$1-\mathcal{O}\left( n^{-\left( 2+\epsilon \right)} \right)$
for any $\epsilon\in \left( 0,2 \right]$.

For the last inequality, note that $z=( e^{\iota \theta_1},\cdots,e^{\iota \theta_n}
)^{\top}$, $e^{\iota k \left( \theta_i-\theta_j
  \right)}=z_i^k\overline{z_j^k}$, and $W_{ij}^{\left( k
  \right)}=u^{\left( k \right)}_i\overline{u_j^{\left( k
    \right)}}$, and note that $\left| z_i^k \right|=1$ for
all $1\leq i\leq n$ and $1\leq k\leq k_{\mathrm{max}}$. We have
\begin{equation*}
  \begin{aligned}
    &\left| W_{ij}^{\left( k \right)}-e^{\iota k \left( \theta_i-\theta_j \right)} \right|=\left|  u^{\left( k \right)}_i\overline{u_j^{\left( k
    \right)}} - z_i^k\overline{z_j^k}\right|\\
    &\leq \left|
u_i^{\left( k \right)} \right|\cdot \left| u_j^{\left( k
  \right)}-z_j^k \right|+\left| z_j^k \right|\cdot \left| u_i^{\left( k \right)}-z_i^k\right|\\
    &\stackrel{\textrm{(Lemma~1)}}{\leq} \left(
      1+C_0\sigma\sqrt{\frac{\log n}{n}}+1\right)\cdot
    C_0\sigma\sqrt{\frac{\log n}{n}}\\
    &<\left( 2+C_0c_0 \right)C_0\sigma \sqrt{\frac{\log n}{n}}
  \end{aligned}
\end{equation*}
where in the last inequality we used the assumption
$\sigma<c_0\sqrt{n/\log n}$.
\end{proof}

\subsection{Proof of Lemma~2}
\label{sec:proof-lemma-2}

\begin{proof}[Proof of Lemma~2]
The proof starts with some
elementary observations for the Dirichlet kernel
$\mathrm{Dir}_m:\left[ 0,2\pi \right]\rightarrow\mathbb{R}$,
defined as
\begin{equation}
  \label{eq:dir-kernel-def}
  \mathrm{Dir}_m \left( x \right)=\sum_{k=-m}^m e^{\iota
    kx}=\frac{\sin \left[ \left( m+1/2 \right)x
    \right]}{\sin \left( x/2 \right)}.
\end{equation}
Note the following (cf.~Figure~\ref{fig:dir-kernel}):
\begin{enumerate}[(1)]
\item\label{item:4} $\left|\mathrm{Dir}_m \left( x
  \right)\right|$ is upper bounded by $1/\sin \left( x/2 \right)$;
\item\label{item:5} $\left|\mathrm{Dir}_m \left( x \right)\right|$
  vanishes at $2\pi\ell/\left(2m+1\right)$, for $\ell
  \in \left[ 2m \right]$;
\item\label{item:6} A unique local maximum exists between
  each pair of consecutive zeros on $\mathbb{R}/2\pi$.
\end{enumerate}
Let $\theta_{*}$ be the local maximizer attaining the highest
``side lobe'' of $\left| \mathrm{Dir}_m \left( x \right)
\right|$ between $2\pi/\left( 2m+1 \right)$ and $4\pi/\left(
  2m+1 \right)$ in Figure~\ref{fig:dir-kernel}. When
$\phi\in\left[ \theta_{*},2\pi-\theta_{*}\right]$, by
Lemma~1, the periodogram
$\left|\mathrm{Re}\left\{\sum_{k=1}^{k_{\mathrm{max}}}
    W^{\left( k \right)}_{ij}e^{-\iota
      k\phi}\right\}\right|$ will not exceed
  \begin{align}
    &\frac{1}{2}\left[\sin \left( \theta_{*}/2 \right) \right]^{-1} -\frac{1}{2} + 2C_2
  k_{\mathrm{max}}\sigma\sqrt{\log n/n}\notag\\
    &\leq \frac{1}{2}\left[\sin \left( \frac{\pi}{ 2m+1 }
      \right)\right]^{-1} -\frac{1}{2} + 2C_2
  k_{\mathrm{max}}\sigma\sqrt{\log n / n}. \label{eq:part-1}
  \end{align}
On the other hand, again by
Lemma~1, the periodogram
$\left|\mathrm{Re}\left\{\sum_{k=1}^{k_{\mathrm{max}}}
    W^{\left( k \right)}_{ij}e^{-\iota
      k\phi}\right\}\right|$ stays above
\begin{align}
  &\frac{1}{2}\left| \mathrm{Dir}_m \left( 0 \right)-1\right|-2C_2
  k_{\mathrm{max}}\sigma\sqrt{\log n/n} \notag \\
   &=m-2C_2
  k_{\mathrm{max}}\sigma\sqrt{\log n/n}. \label{eq:part-2}
\end{align}
Therefore, as long as the upper bound \eqref{eq:part-1} is
no greater than the lower bound \eqref{eq:part-2}, which one
can check is satisfied if condition (19) in the state of the
lemma holds, i.e., if
\begin{equation*}
  \left[2k_{\mathrm{max}}\sin \left(
        \frac{\pi}{2k_{\mathrm{max}}+1} \right)
    \right]^{-1}+4C_2\sigma\sqrt{\log n / n}<1
\end{equation*}
then the peak location of the periodogram
$\left|\mathrm{Re}\left\{\sum_{k=1}^{k_{\mathrm{max}}}
    W^{\left( k \right)}_{ij}e^{-\iota
      k\phi}\right\}\right|$
can occur nowhere other than within $\left[ 0,\theta_{*}
\right]\cup\left[ 2\pi-\theta_{*},2\pi \right]$, which gives
the conclusion
\begin{equation*}
  \left| \hat{\theta}_{ij}-\left( \theta_i-\theta_j \right)
  \right|\leq \theta_{*}<\frac{4\pi}{2m+1}
\end{equation*}
with $m=k_{\mathrm{max}}$. This completes the proof.

\begin{figure}[htbp]
\centering
\includegraphics[width=0.5\textwidth]{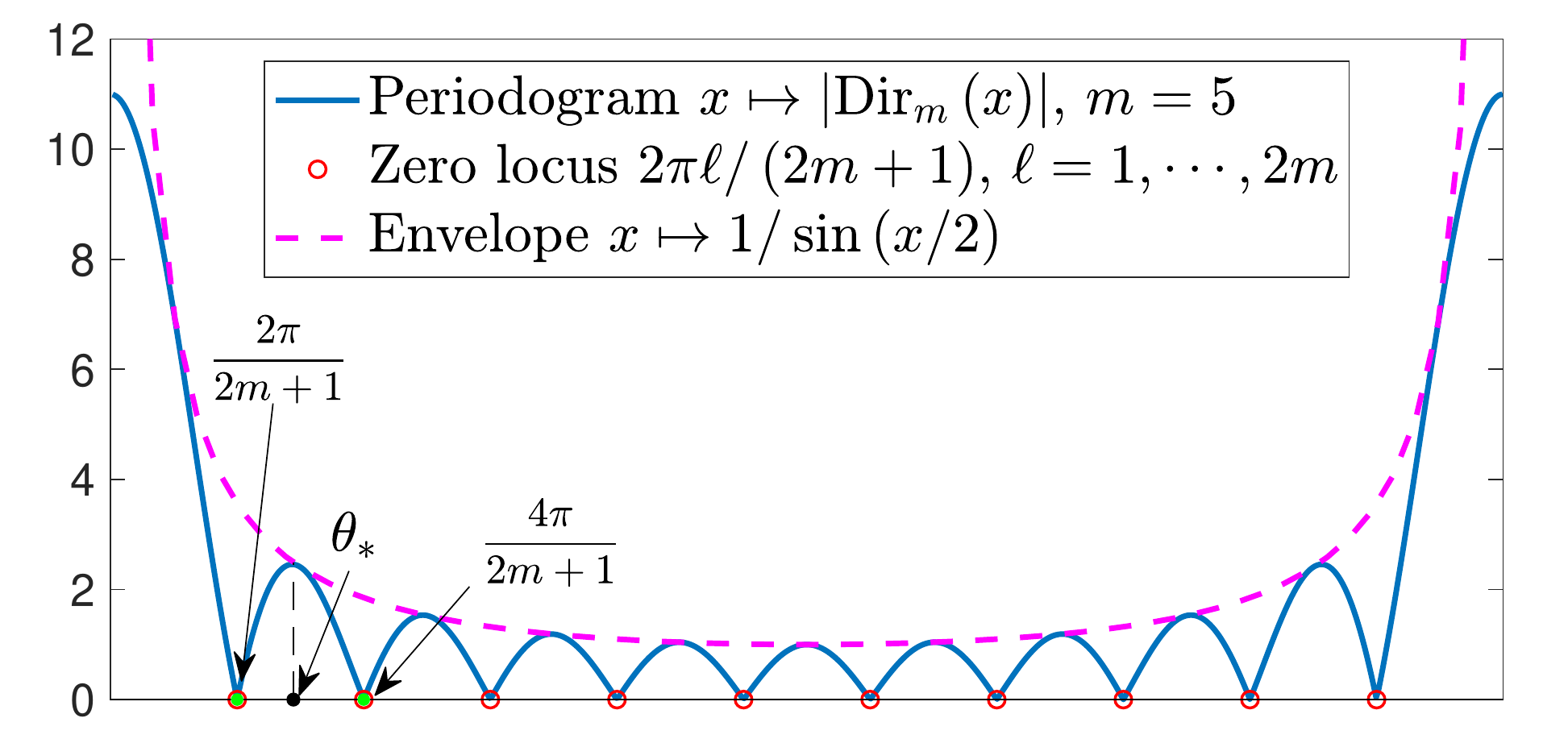}
\caption{Illustration of the periodogram of a Dirichlet
  kernel function with $m=5$ on $\left[ 0,2\pi \right]$, for the proof
  of Lemma~2.}
\label{fig:dir-kernel}
\end{figure}
\end{proof}

\subsection{Proof of Theorem~2}
\label{sec:proof-theorem-2}

\begin{proof}[Proof of Theorem~2]
First, we note that the second part of the theorem about
$\hat{x}$ follows directly from Proposition~1 of
\cite{LYM2017}, as in the proof of Lemma~8 of \cite{ZB2018}.

Assuming for the moment that the key assumption in Lemma~2 is
satisfied, namely, $n$ and $k_{\mathrm{max}}$ have been
chosen such that
\begin{equation}
\label{eq:copy-key-gap-condition}
  \frac{1}{\displaystyle 2k_{\mathrm{max}}\sin \left( \frac{\pi}{2k_{\mathrm{max}}+1} \right)}+4C_2\sigma\sqrt{\frac{\log n}{n}}<1.
\end{equation}
  With a union bound over each of the $\mathcal{O}\left( n^2
  \right)$ estimated relative phases $\hat{\theta}_{ij}$
  obtained at the end of the Step 2 of Algorithm~1, with
  probability at least $1-\mathcal{O} \left( n^2\cdot
    n^{-\left( 2+\epsilon \right)}
  \right)=1-\mathcal{O}\left( n^{-\epsilon} \right)$ we have
  for all $\left( i,j \right)\in E$
  \begin{equation*}
    \left| \hat{\theta}_{ij}-\left( \theta_i-\theta_j
      \right) \right|\leq
    \frac{4\pi}{2k_{\mathrm{max}}+1}
  \end{equation*}
and thus
  \begin{equation}
   \label{eq:entry-aux}
   \begin{aligned}
     \left| \widehat{H}_{ij} -z_i\overline{z}_j \right| &=
    \left| e^{\iota \hat{\theta}_{ij}}-e^{\iota \left(
          \theta_i-\theta_j \right)} \right|\\
     &\leq \left| \hat{\theta}_{ij}-\left( \theta_i-\theta_j
      \right) \right|\leq
    \frac{4\pi}{2k_{\mathrm{max}}+1}.
   \end{aligned}
  \end{equation}
Therefore,
\begin{equation}
\label{eq:op-norm-bound}
  \left\| \widehat{H}-zz^{*} \right\|_2\leq \left\|
    \widehat{H}-zz^{*} \right\|_{\mathrm{Frob}}\leq \frac{4\pi n}{2k_{\mathrm{max}}+1}
\end{equation}
where the last equality follows from
bounding each entry of $H-zz^{*}$ individually using the
rightmost term in \eqref{eq:entry-aux}. (Note that by doing
so we do not need any information on the randomness of
$H-zz^{*}$.) By the Davis--Kahan $\sin \Theta$ Theorem in
Lemma~11 of \cite{ZB2018}, as long as $n>\left\|
  \widehat{H}_{ij}-zz^{*} \right\|_2$, which we know from
\eqref{eq:op-norm-bound} that can be guaranteed if
$k_{\mathrm{max}}>2\pi-1/5\approx 5.7832$, the angle $\theta
\left( \hat{u},z \right)$ between $\hat{u}$ and $z$ satisfies
\begin{equation*}
  \begin{aligned}
    \sin\theta \left( \hat{u},z \right)&\leq \frac{\left\| \widehat{H}-zz^{*} \right\|_2}{n-\left\| \widehat{H}-zz^{*} \right\|_2}\leq \frac{\displaystyle
      \frac{4\pi n}{2k_{\mathrm{max}}+1}}{\displaystyle n-\frac{4\pi
        n}{2k_{\mathrm{max}}+1}}\\
     &=\frac{4\pi}{2k_{\mathrm{max}}+1-4\pi}<\frac{400\sqrt{2}\,\pi}{k_{\mathrm{max}}}
  \end{aligned}
\end{equation*}
where in the last inequality we used the fact that $\left(
  2-0.01 \right)k_{\mathrm{max}}\geq 4\pi-1$ for all
$k_{\mathrm{max}}\geq 6$. Therefore, setting $C_3:=\left(
  400\sqrt{2}\,\pi \right)^2$, we have
\begin{equation*}
  \begin{aligned}
    \frac{\left| \hat{u}^{*}z \right|}{\left\| \hat{u}
      \right\|_2 \left\| z \right\|_2}&=\left|\cos\theta \left(
      \hat{u},z \right)\right|\\
      &\geq \cos^2 \theta \left(
    \hat{u},z \right) = 1-\sin^2 \theta \left( \hat{u},z
  \right)\\
      &\geq 1-\frac{C_3}{k_{\mathrm{max}}^2}.
  \end{aligned}
\end{equation*}

Now we seek lower bound for $n$ and $k_{\mathrm{max}}$ that
satisfies \eqref{eq:copy-key-gap-condition} under the 
condition $\sigma<c_0\sqrt{n/\log n}$ imposed in
Lemma~1. Obviously, \eqref{eq:copy-key-gap-condition} is
satisfied if
\begin{equation}
\label{eq:aux-2}
  2k_{\mathrm{max}}\sin \left(
      \frac{\pi}{2k_{\mathrm{max}}+1}
    \right)>\frac{1}{\displaystyle 1 - 4C_2\sigma\sqrt{\log
        n / n}}.
\end{equation}
Using the elementary inequality
\cite{Kroopnick1997}
\begin{equation*}
  \sin x > \frac{x}{\sqrt{1+x^2}},\qquad\forall x>0.
\end{equation*}
we know that a sufficient condition for \eqref{eq:aux-2} to
hold is
\begin{equation}
  \frac{\displaystyle
    2k_{\mathrm{max}}\cdot \frac{\pi}{2k_{\mathrm{max}}+1}}{\displaystyle
  \sqrt{1+\frac{\pi^2}{\left( 2k_{\mathrm{max}}+1
      \right)^2}}}>\frac{1}{\displaystyle 1 - 4C_2\sigma\sqrt{\log
        n / n}},
\end{equation}
which is further equivalent to
\begin{equation}
  \label{eq:aux-4}
  \frac{2k_{\mathrm{max}}\pi}{\sqrt{\left( 2k_{\mathrm{max}}+1 \right)^2+\pi^2}}>\frac{1}{\displaystyle 1 - 4C_2\sigma\sqrt{\log n / n}}.
\end{equation}
Note that for all $k_{\mathrm{max}}\geq 2$ we have
$2k_{\mathrm{max}}+1>\pi$, and thus $\left(
  2k_{\mathrm{max}}+1 \right)^2+\pi^2<2 \left(
  2k_{\mathrm{max}}+1 \right)^2$. Therefore, a sufficient
condition for \eqref{eq:aux-4} to hold is
\begin{equation*}
  \begin{aligned}
    &\frac{\sqrt{2}\,\pi
    k_{\mathrm{max}}}{2k_{\mathrm{max}}+1}>\frac{1}{\displaystyle
    1 - 4C_2\sigma\sqrt{\log n /
      n}}\quad\Leftrightarrow\quad\\
    &k_{\mathrm{max}}>\frac{1}{\sqrt{2}\,\pi \left( 1-4C_2\sigma\sqrt{\log n / n}\right)-2}.
  \end{aligned}
\end{equation*}
\end{proof}

\section{Extra Numerical Results}
\label{sec:extra_results}

\begin{figure}
\captionsetup[subfloat]{farskip=2pt,captionskip=1pt}
\begin{center}
\subfloat[PPE-SPC]{
\includegraphics[ width =  0.3\columnwidth]{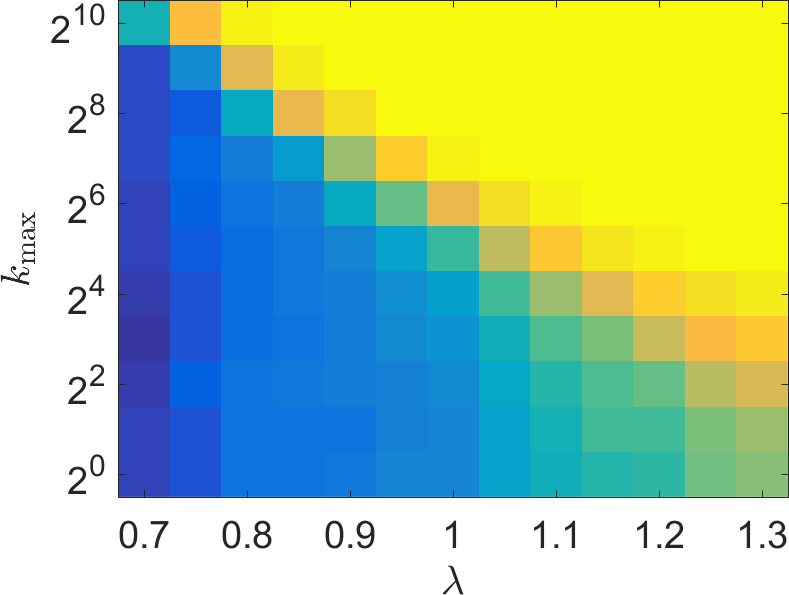}
\label{fig:gau_mc_spec1_n100}
}
\subfloat[PPE-SPC$^3$]{
\includegraphics[ width =  0.3\columnwidth]{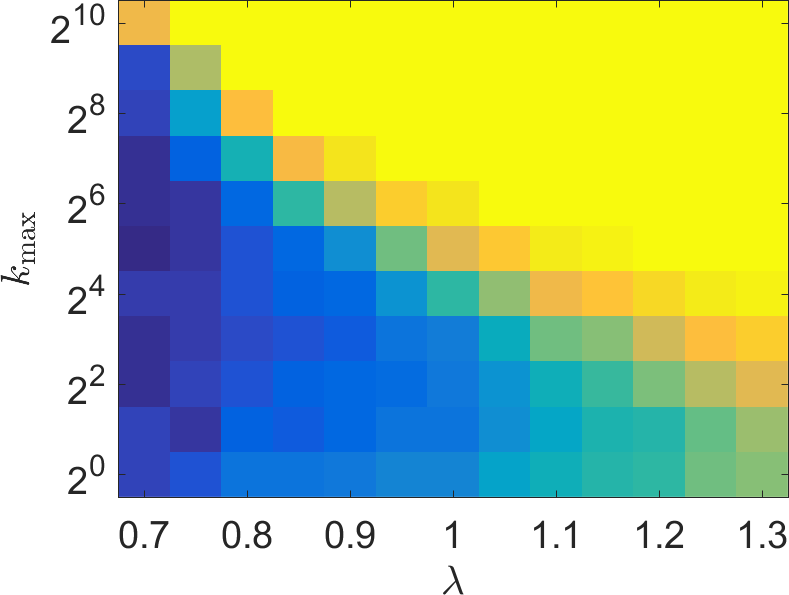}
\label{fig:gau_mc_spec2_n100}
}
\subfloat[AMP]{
\includegraphics[ width = 0.3\columnwidth]{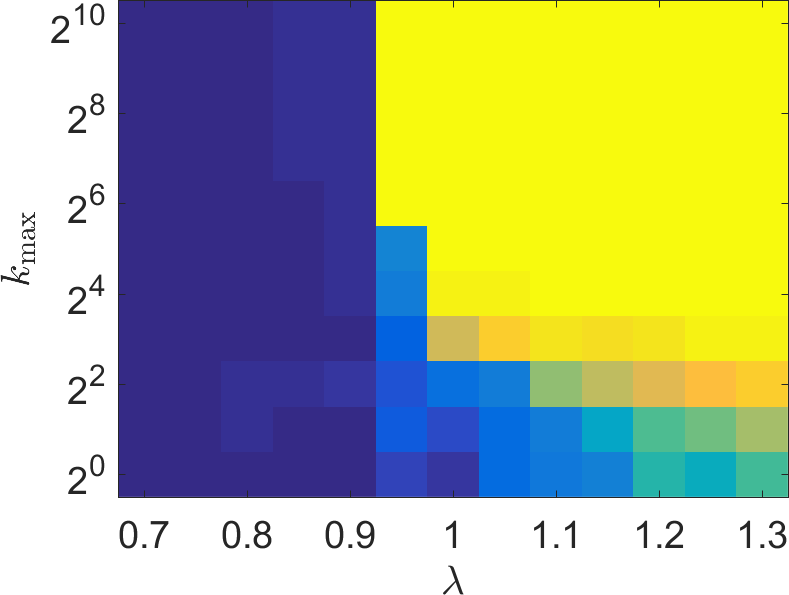}
\label{fig:gau_mc_AMP_n100}
}
\includegraphics[height=0.23\columnwidth]{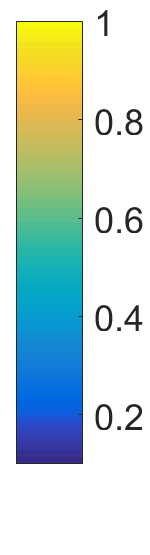}
\end{center}
\caption{$\Unitary\left(1\right)$ synchronization under
  Gaussian noise model with $\sigma=\sqrt{n}/\lambda$ for $n = 100$ vertices. Every data point is the median over 20 trials. }
\label{fig:n100_mc_gau}
\end{figure}

We consider the incomplete graph structure with $n = 100$
vertices under Erd\H{o}s--Renyi graph model and the edge
connection probability $p = 0.23$ for the following
experiments. Figure~\ref{fig:n100_mc_gau} shows that
Algorithm~1 (PPE-SPC) is also robust for incomplete graphs.

\begin{figure}
\captionsetup[subfloat]{farskip=2pt,captionskip=1pt}
\begin{center}
\subfloat[PPE-SPC]{
\includegraphics[ width = 0.3\columnwidth]{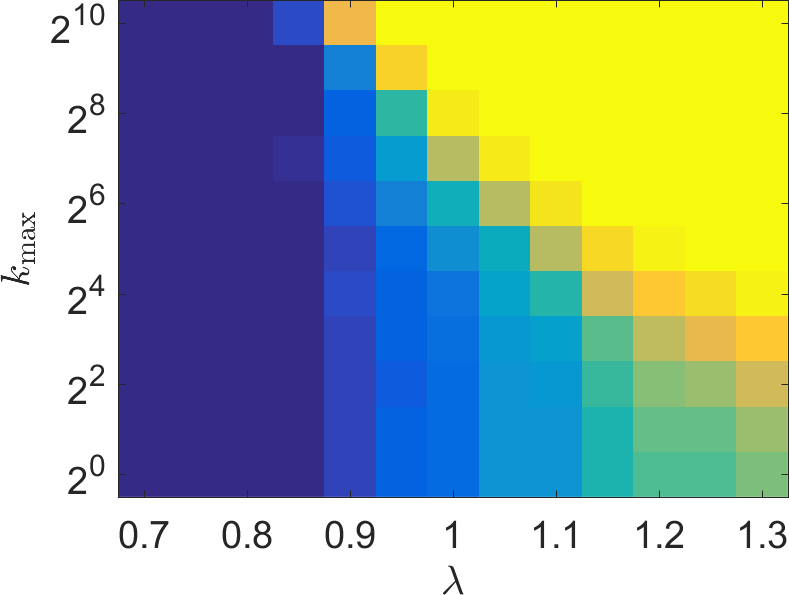}
\label{fig:gau_cg_spec1_n500}
}
\subfloat[PPE-SPC$^{3}$]{
\includegraphics[ width =  0.15\textwidth]{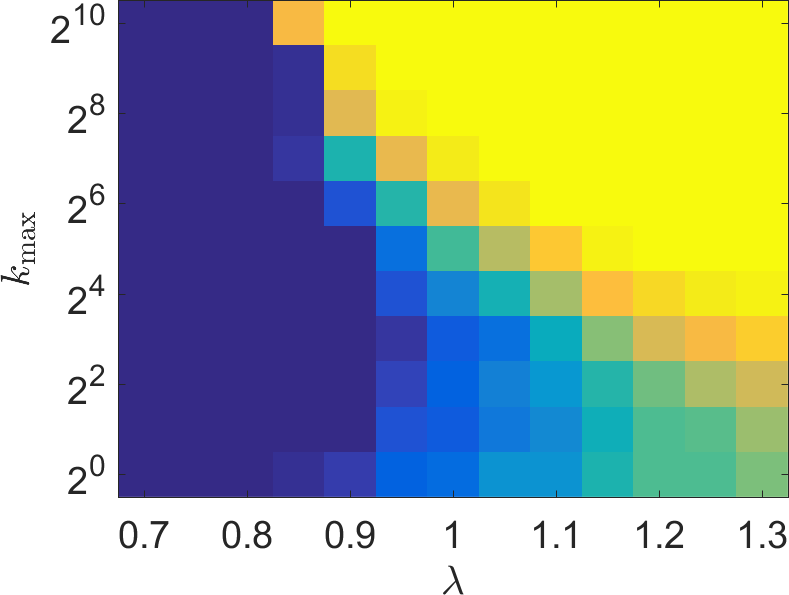}
\label{fig:gau_cg_spec2_n500}
}
\subfloat[AMP]{
\includegraphics[ width = 0.3\columnwidth]{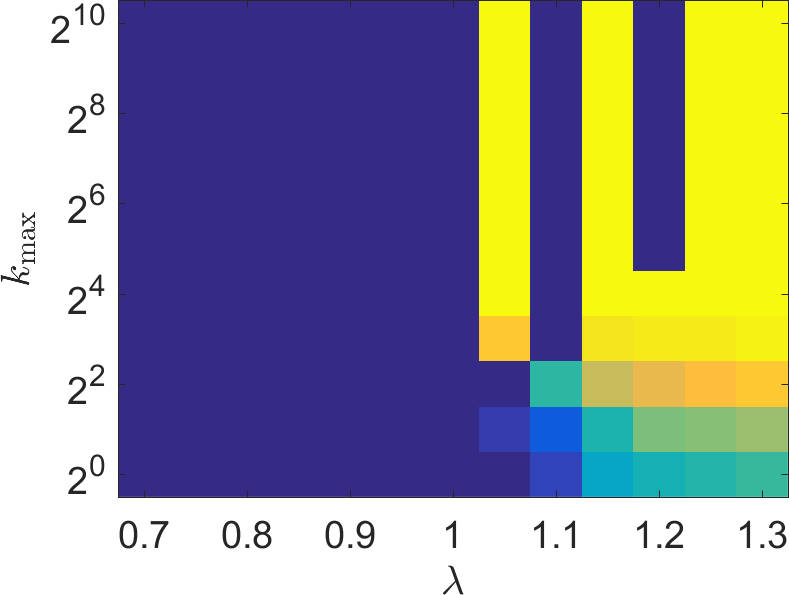}
\label{fig:gau_cg_AMP_n500}
}
\includegraphics[height=0.23\columnwidth]{./figures/color_bar2}
\end{center}
\caption{Correlation value for $\Unitary\left(1\right)$ synchronization under Gaussian noise model for $n = 500$ vertices.  }
\label{fig:n500_cg_gau}
\end{figure}

\begin{figure}
\captionsetup[subfloat]{farskip=2pt,captionskip=1pt}
\begin{center}
\subfloat[PPE-SPC]{
\includegraphics[ width =  0.3\columnwidth]{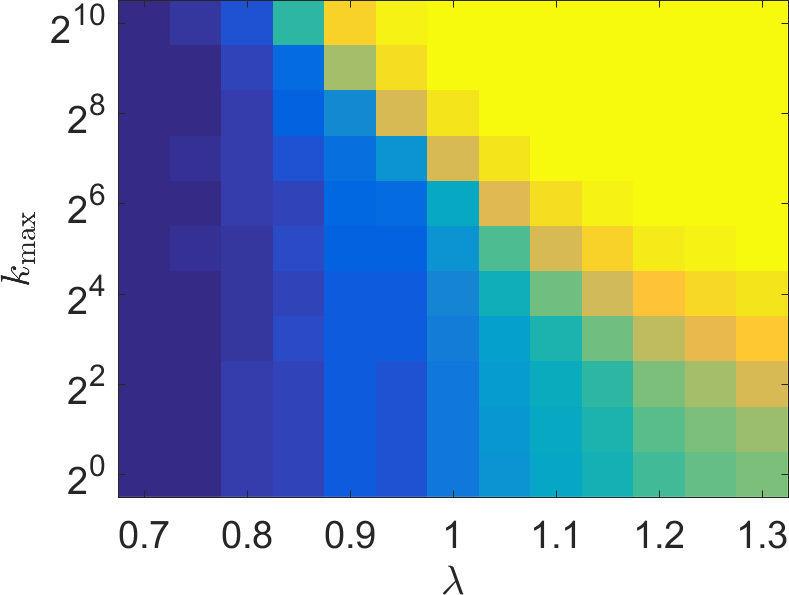}
\label{fig:rc_cg_spec1_n500}
}
\subfloat[PPE-SPC$^3$]{
\includegraphics[ width =  0.3\columnwidth]{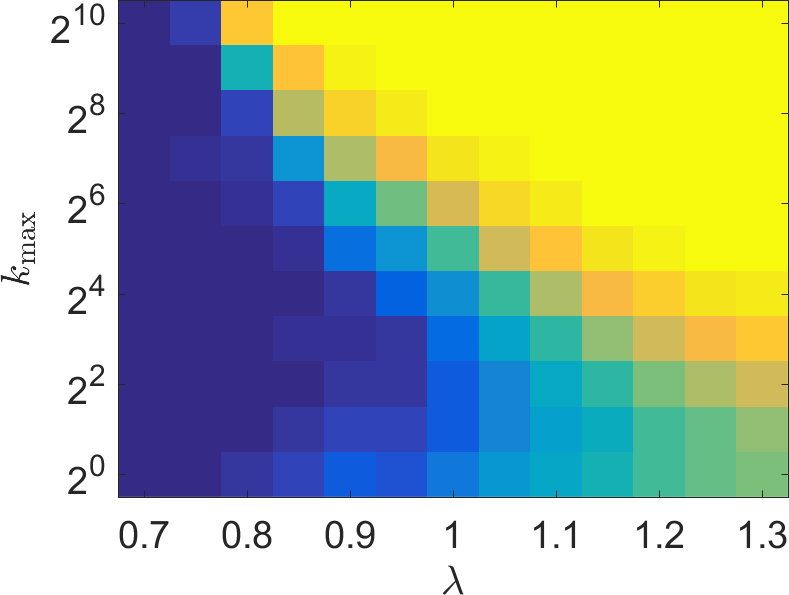}
\label{fig:rc_cg_spec2_n500}
}
\subfloat[AMP]{
\includegraphics[ width = 0.3\columnwidth]{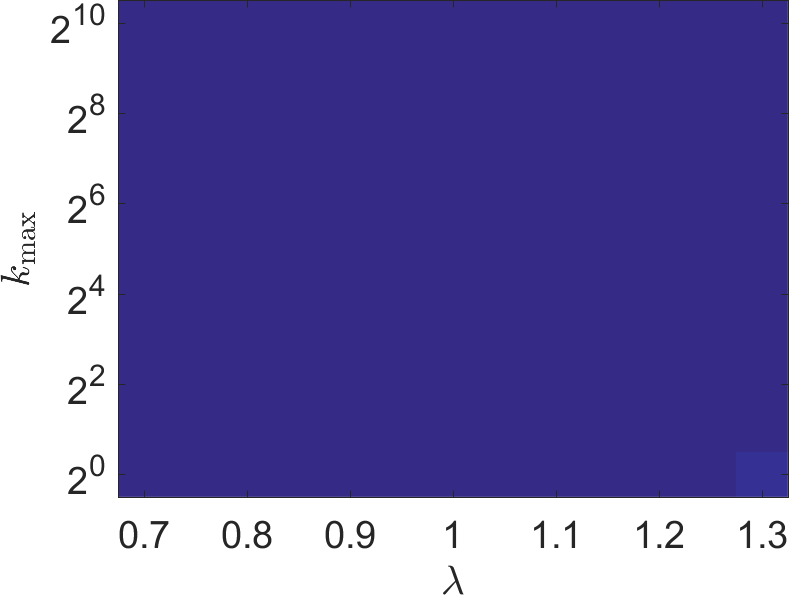}
\label{fig:rc_cg_AMP_n500}
}
\includegraphics[height=0.23\columnwidth]{./figures/color_bar2}
\end{center}
\caption{Correlation value for $\Unitary\left(1\right)$ synchronization under random corruption model with $r = \frac{\lambda}{\sqrt{n}}$ for $n = 500$ vertices and fully connected graph.}
\label{fig:n500_cg_rc}
\end{figure}

Figures~\ref{fig:n500_cg_gau} and~\ref{fig:n500_cg_rc} show the performance of our PPE-SPC and its variant PPE-SPC$^3$ on complete graph with $n = 500$ vertices. 

\bibliography{ref}
\bibliographystyle{icml2019}

\end{document}